\documentclass[fleqn,10pt]{wlscirep}
\usepackage[utf8]{inputenc}
\usepackage[T1]{fontenc}
\usepackage[square,super,compress,comma,numbers]{natbib}

\usepackage{graphicx}
\usepackage{bm}

\usepackage{mathptmx}
\usepackage{etoolbox}

\usepackage{cancel}

\usepackage{blindtext}

\usepackage[style=base]{subcaption}

\usepackage{float}

\usepackage{color}
\usepackage[noabbrev]{cleveref}

\makeatletter
\def\@email#1#2{%
 \endgroup
 \patchcmd{\titleblock@produce}
  {\frontmatter@RRAPformat}
  {\frontmatter@RRAPformat{\produce@RRAP{*#1\href{mailto:#2}{#2}}}\frontmatter@RRAPformat}
  {}{}
}%
\makeatother

\title{Statistical machine learning tools for probabilistic closures of turbulence models}

\author[1,2,*]{J. Domingues Lemos}
\author[3]{F. P. Santos}
\affil[1]{École Normale Supérieure de Lyon, Laboratoire de Physique, 46 allée d’Italie 69007 Lyon, France.}
\affil[2]{{Federal University of Rio de Janeiro}, {Systems Engineering and Computer Science Program}, {{Rio de Janeiro}, {21941-909}, {Rio de Janeiro}, {Brazil}}}
\affil[3]{%
Laboratório Nacional de Computação Científica, Av. Getulio Vargas, 333 - Quitandinha, Petrópolis - RJ, 25651-075}

\affil[*]{julia.domingues\_lemos@ens-lyon.fr}

\begin{abstract}
{Turbulent flow remains a challenging subject, despite extensive efforts to find analytical descriptions. Modeling small scales of motion is crucial for saving time and resources in numerical simulations, particularly in industrial applications. Here we attempt to model small scales of motion by creating closures for a Shell model of turbulence, more specifically for Sabra. Shell models are infinite dimensional dynamical systems that retain most key properties of Navier-Stokes equation, such as energy cascade and intermittency, while being computationally treatable. To account for Sabra's intermittent fluctuations we employ a set of scaling relations that recover a hidden symmetry and leaves us with universal statistics across the inertial range. On data from these rescaled variables we then adapt and apply two machine learning tools, a variational auto-encoder and sparse identification of non-linear dynamics. We estimate the data's densities and dynamics in order to then generate new instances of data. Given a mid-inertial range cutoff scale, we evolve reduced models in time, resolving only scales of motion larger than the cutoff scale, closing the reduced model with data generated by the machine learning tools. We compare statistics of our reduced models against a Sabra's fully resolved simulation to evaluate each closure's performances. Our results show improvement regarding previous work, and all our closures are probabilistic and cutoff-independent.}
\end{abstract}


\begin{document}

\flushbottom
\maketitle
%
%
\thispagestyle{empty}

\noindent \textbf{The modeling of small scales of motion is of extreme importance to the understanding of turbulent flows, but also to make numerical simulations feasible for most realistic applications. We elaborate closures for turbulence models in the form of random variables that present adequate probability density functions, or in the form of a stochastic process that satisfies an approximate dynamic. Furthermore, we do so with the aid of a set of spatio-temporal rescaling relations, which allows us to deal with intermittent velocity fluctuations, and machine learning tools, which we tailored to accomplish the required tasks. Our results show that the closures proposed here are an improvement on works under the same theoretical framework, while, at the same time, complying to the recent advances that emphasize the probabilistic nature of turbulent flows.}

\section{Introduction}

The Navier-Stokes equations govern fluid flows and serve as the foundation of fluid dynamics. In this context, closure problems arise when the number of variables exceeds the number of available equations. Specifically, we address closure problems related to subgrid modeling, which aims to describe the small-scale motions in turbulent flows \cite{clos_eyink}. Such modeling becomes necessary due to the phenomenon of energy cascade in turbulent flows, where energy injected at large motion scales cascades through the inertial range until it dissipates as heat at sufficiently small scales due to viscosity \cite{frisch}. 

Many closures were written to the Navier-Stokes equations in approaches such as Large Eddy Simulation (LES) or Reynolds-Average Navier-Stokes (RANS) \cite{reynoldsaverage,les_1,pope2000turbulent}, and those frequently make use of phenomenological predictions and/or calibrated models, which is sufficient for many industrial applications \cite{nudging,clos_eqfree}. We would like, however, to be able to write a model for small scales of motion solely from data that provides an accurate description of the flow for all scales of motion larger than a certain cutoff scale. Moreover, we would like to write closures that are probabilistic to accommodate the recent advances in works of spontaneous stochasticity that suggest that the solution of the Navier-Stokes equations in turbulent regimes evolve as a stochastic process \cite{Lorenz69,leithkraich,ruelle,eyinknoise,massimum,Palmer,Mailybaev_2016,Mailybaev_2017,RTbifbof,spont1,bandak2024}.

To better understand the statistical and qualitative behavior of closure problems in turbulent flows, instead of tackling Navier-Stokes directly, we will make use of a shell model \cite{gled,goy,lucashell}. Shell models are infinite dimensional dynamical systems that describe fluid motion in a scale-by-scale manner. They are both analytically and computationally more tractable while preserving critical properties of the original Navier-Stokes equations, including the energy cascade, intermittency, and inviscid invariants.

In the context of shell models, more specifically for a shell model called Sabra \cite{lvov}, closure problems are reduced to finding expressions for the missing variables, which we will call closure variables, present in the non-linear coupling of scales of motion. One significant difficulty in closure problems is the presence of intermittency, which makes the statistics of velocity fluctuations non-universal and indicates that a suitable closure should be able to recover multifractal properties from small scales\cite{lucashell}. There is a set of variables that can be recovered from Sabra variables that do present universal statistics\cite{benzi,eyink2003}, they are referred to as Kolmogorov multipliers \cite{kolmogorov_1962}. In \cite{Biferale_2017}, the authors propose a closure model based on these statistics. However, this approach is constrained by the non-universality of multitime statistics for the multipliers. 

Our approach to handle intermittency relies on employing a set of scaling relations that defines local times based on a reference scale and rescales each scale of motion accordingly \cite{alexei,Mailybaev2022HiddenSI}. This recovers a hidden scale invariance of statistics across the inertial range, which is also valid for multitime statistics. We are referring to this set of rescaled variables as the Hidden Symmetry framework. This set of scaling relations was first proposed in \cite{alexei} and was used in \cite{euzinha} to write closures that use Gaussian Mixture Models (GMM) \cite{bishop} to generate the missing closure variables, and it succeeded in producing probabilistic, time-correlated, cutoff-independent closures.

Other recent works in closure problems for shell models include \cite{lstmsabra,sabrann}, which used different types of neural networks, namely a Long Short-Term Memory network and a solver-in-the-loop approach, to generate closure variables at each time step. These works have extremely accurate results and are significant contributions that further our understanding of this problem. However, in general, when a machine learning tool is trained directly on Sabra variables, it is expected that the tool will need to be retrained when faced with a change in cutoff scale or in forcing.

In an attempt to improve on the results reported in \cite{euzinha}, which used the Hidden Symmetry framework and GMM,  we choose here two other machine learning tools. The first is a Variational Auto-Encoder (VAE) \cite{vae}, a neural network with a variational layer that learns the underlying distribution of data and generates new instances of data on the same distribution. Our second tool is called Sparse Identification of Nonlinear Dynamics (SINDy) \cite{sindy}, which works on the assumption that one has data, perhaps trajectories of some quantities evolving in time, that are the solution of a dynamical system, and SINDy estimates such system. It does so by estimating which functions and with which coefficients are present on the dynamic.

We use the closure variables generated by our machine learning tools to evolve reduced models, consisting of only the large scales of motion up to the cutoff scale, in time. We then evaluate how well the closure is performing by comparing statistics and observables of the reduced models against the ones collected from a fully resolved Sabra simulation. In section \ref{sec:shells} we describe Sabra, the rescaled variables and reduced models. In section \ref{sec:tools} we discuss the machine learning tools we chose and in \ref{sec:results} we report the results. In section \ref{sec:disc} we discuss and interpret the results, as well as indicate future works.

\section{Shell models and closure problems }\label{sec:shells}

Shell models are infinite-dimensional dynamical systems that provide a simplified yet effective representation of velocity fluctuations, retaining the core characteristics of the Navier-Stokes equations \cite{lucashell,frisch}. The model we are using is called Sabra \cite{lvov}, and can be written as
\begin{align}
    \frac{du_n}{dt} = i\big(k_{n+1}u_{n+2}u_{n+1}^* - \frac{1}{2}k_n u_{n+1}u_{n-1}^* + \frac{1}{2} k_{n-1}u_{n-1}u_{n-2}\big) - \nu k_n^2u_n + f_n. \label{eqn:sabra}
\end{align}
where $u_n \in \mathbb{C}$ describes velocity fluctuations at shell $n \in \mathbb{N}$, $\nu$ is the kinematic viscosity and $f_n$ is the forcing term only active via non-zero constant in the first component, i.e., the integral scale $L=1/k_0$. Wavenumbers are discretized in a geometric progression as $|\mathbf{k}| = k_n = k_0 \lambda ^n$ with, typically, $\lambda=2$ and $k_0 = 1$. By defining the characteristic velocity $U = \sqrt{|f_1|/k_0}$ we can also define the Reynolds number $\mathrm{Re}=UL/\nu$. The wavenumbers corresponding to the forcing range are of the order $k_n/k_0 \sim 1$, while the inertial range, where energy in injected and transferred to smaller scales, contemplates the wave numbers $1 \ll k_n/k_0 \ll \mathrm{Re}^{3/4}$ \cite{frisch1991prediction,mailybaev2023hidden}. Energy is then transferred until the dissipation range, of wavenumbers $k/k_0 \gtrsim \mathrm{Re}^{3/4}$.

Equations \eqref{eqn:sabra} give a scale-by-scale description of velocity fluctuations. The nonlinear terms present a coupling of scales that require boundary conditions for $u_0$ and $u_1$, which are always set as $u_0=u_1=0$. This indicates that there is no motion happening in scales larger than the integral scale. When computing a simulation of Sabra, we need to choose the amount $s$ of scales of motion to be resolved, and $s$ needs to be large enough to cover forcing, inertial and dissipation range, in order to achieve a physically correct simulation. If that is the case, then we can set $u_{s+1}=u_{s+2}=0$ and equations \eqref{eqn:sabra} are closed. Resolving enough scales in a shell model is roughly equivalent to having a computational grid with spacing smaller than the dissipative scale in a direct numerical simulation of homogeneous isotropic turbulent flow governed by the Navier-Stokes equation.

When $s$ is not large enough to cover all three ranges, say, if it is only enough to reach mid-inertial range, we are faced with the task of finding a way to model $u_{s+1}$ and $u_{s+2}$ in order to close equations \eqref{eqn:sabra}. Since Sabra's shell velocities do not have universal statistics along the inertial range due to intermittency \cite{lucashell}, any modeling done for one specific value of $s$ will not suffice in closing our system if a different value of $s$ is required. 


We follow the set of spatio-temporal rescaling relations first presented in \cite{alexei}, defined by first choosing a reference shell $m$ in the inertial range, then writing a local time
\begin{gather}
    T_m(t) = \left(k_0^2 U^2 + \sum_{n <m} k_n^2|u_n|^2\right)^{-1/2},
\end{gather}

\noindent and then defining the implicit change of variables
\begin{gather}
    \tau = \int_0^t \frac{dt'}{T_m(t')}, \label{eqn:taut}\\
    \mathcal{U}_N = k_m T_m(t) u_{N+m}(t)     \label{eqn:change}
\end{gather}

As demonstrated in \cite{euzinha}, the Sabra system from equation \eqref{eqn:sabra} can be reformulated in its rescaled version, resulting in the complete (forced and viscous) system as follows:
\begin{align}
    \frac{d\mathcal{U}_N}{d\tau} = i(k_{N+1}\mathcal{U}_{N+2}\mathcal{U}_{N+1}^* -\frac{1}{2}k_{N} \mathcal{U}_{N+1}\mathcal{U}_{N-1}^* + \frac{1}{2}k_{N-1}\mathcal{U}_{N-1}\mathcal{U}_{N-2}) +\left( \xi_{total} - \nu k_{N+m}^2T_m \right) \mathcal{U}_{N} + T_m^2 k_m f_{N+m}, \label{eqn:system6}
\end{align}
where
\begin{gather} 
    \xi_{total} = \xi + \xi_{\nu} + \xi_f,  \label{eqn:dtmdt} \\
    \xi = \sum_{N<0} k_{N}^3 \operatorname{Im} \Big{(}  2 \mathcal{U}_{N}^*  \mathcal{U}_{N+1}^* \mathcal{U}_{N+2} - \frac{1}{2} \mathcal{U}_{N-1}^* \mathcal{U}_{N}^* \mathcal{U}_{N+1} - \frac{1}{4} \mathcal{U}_{N-1}^*  \mathcal{U}_{N}  \mathcal{U}_{N-2}^* \Big{)}, \label{eqn:xi}\\
    \xi_{\nu} = \nu T_m k_m^2 \sum_{N<0}  k_{N}^4 |\mathcal{U}_{N}|^2 , \label{eqn:xinu}\\
    \xi_f = - T_m^2 \sum_{N<0}k_{N+m} k_{N} \operatorname{Re}\Big{(} \mathcal{U}_{N}^* f_{N+m} \Big{)},  \label{eqn:xif} \\ 
    T_m = \frac{1}{k_0 U}\left( 1- \sum_{N<0}k_N^2|\mathcal{U}_N|^2 \right)^{1/2}. \label{eqn:Tm_tau}
\end{gather}

\begin{figure}
    \begin{subfigure}{.48\textwidth}
    \subcaption{}
    \includegraphics[scale=0.49]{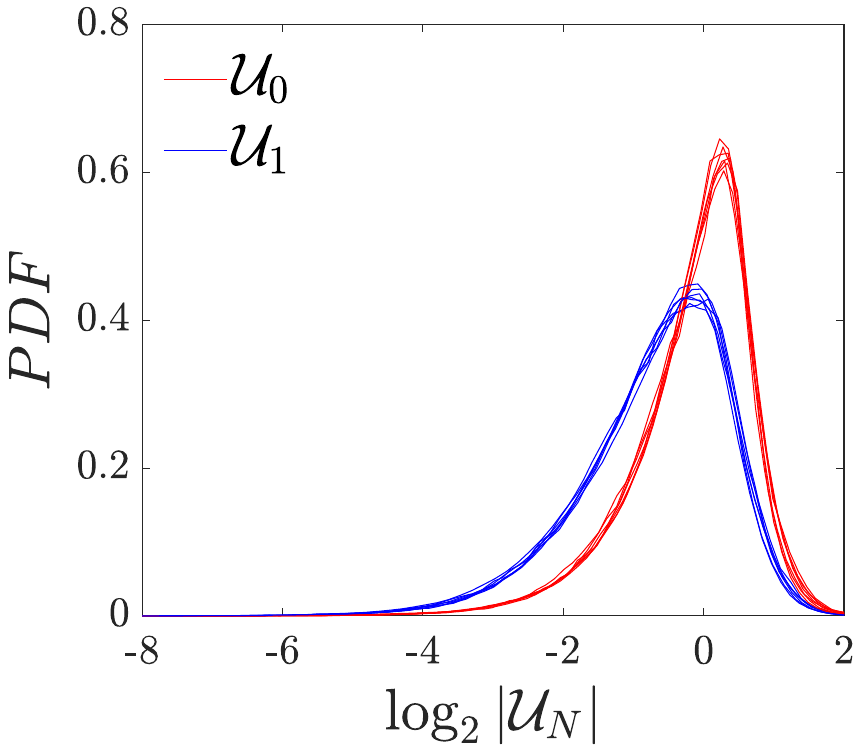}
    \label{fig:absu0u1}    
    \end{subfigure}
    \begin{subfigure}{.48\textwidth}
    \subcaption{}
    \includegraphics[scale=0.56]{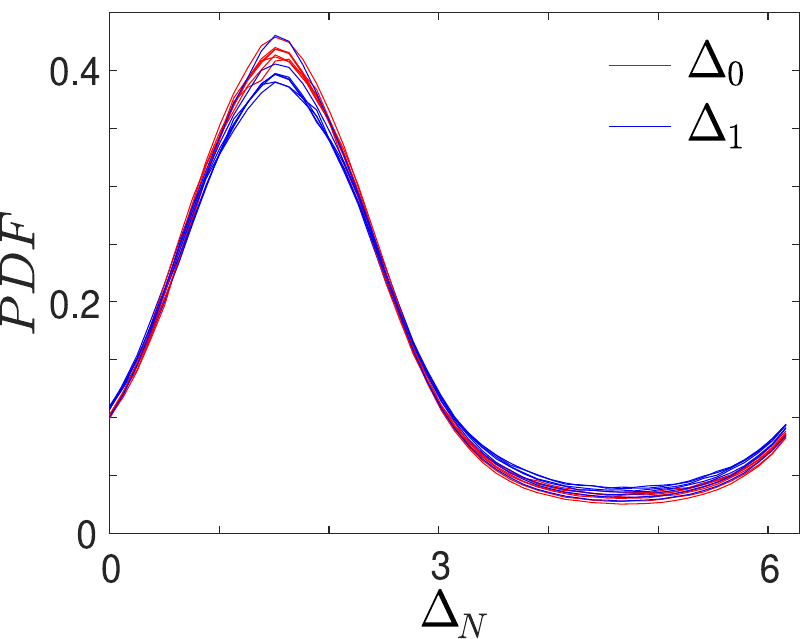}
    \label{fig:delta01}
    \end{subfigure}
    \caption{PDFs of $|\mathcal{U}_0|$ and $|\mathcal{U}_1|$ in \ref{fig:absu0u1} and of multipliers phases in \ref{fig:delta01} for $m = 8,\ldots,14$.}  \label{fig:prim}
\end{figure}

As reported in \cite{euzinha,alexei,mailybaev2022hidden}, the statistics collected in time for the rescaled shell velocities are universal with respect to the reference shell $m$. For different choices of $m$, shell $\mathcal{U}_0$ is always the m-th shell and, despite being different scales of motion for different choices of $m$, they all present the same probability density functions (PDFs). These rescaling relations work to provide a set of variables in which intermittency is not affecting the statistics in the inertial range, but instead has been encoded into the change of variables itself. To see universality, as per figure \ref{fig:prim}, we choose several different reference shells $m = 8,\ldots,14$, corresponding to the several curves, and compute PDFs of $\log_2|\mathcal{U}_0|$ and $\log_2|\mathcal{U}_1|$, as well as their associated multiplier phases given by the phase differences \cite{kolmogorov_1962,frisch}
\begin{gather}
    \Delta_N = \arg(\mathcal{U}_N) - \arg(\mathcal{U}_{N-1}) - \arg(\mathcal{U}_{N-2}). \label{eqn:deltaa}
\end{gather}

The fully resolved simulation of the Sabra system that will serve as ground truth for the training of the machine learning tools, contains $30$ resolved shells, with boundary conditions $u_0 = u_{-1} = 0$, viscosity $\nu = 10^{-8}$, forcing  $f_1 = 1+i$ only active in the first shell and the initial condition comes from a stationary state. We recover rescaled variables by solving equation \eqref{eqn:taut} for $\tau(t)$ and applying the change of variables in \eqref{eqn:change}.

We intend to find a suitable tool in machine learning that can perform a density estimation for the joint PDF of modules of complex velocities $\mathcal{U}_0$ and $\mathcal{U}_1$ and their associated multiplier phases $\Delta_0$ and $\Delta_1$. Such estimation would, ideally, allow us to generate new instances of data for the closure variables, which would then be used to evolve reduced models in time.

The machine learning tools tested here succeeded in many aspects, but presented difficulties in performing a more refined density estimation, such as learning conditional probability densities that would allow us to generate closure variables explicitly conditioned to the system's prehistory, therefore all closures reported here are single-time models. These tools will be discussed in the next section.

\section{Machine learning tools} \label{sec:tools}

In this section we discuss in more detail the tools we chose to perform our density estimations. Firstly, we need a machine learning tool that can handle high-dimensional data. Furthermore, we need to be able to generate new instances of data, so it must be a generative tool, but it needs a stochastic component, to accommodate for the probabilistic nature of turbulent flows. Our first choice is, then, a Variational Auto-Enconder (VAE) \cite{vae}. We do anticipate, however, that VAEs may face challenges in this work, given that they do not behave well working on data that feature a multi-modal density \cite{nonvae}, or a periodic one, and the densities for multipliers phases $\Delta_N$ present either one or the other.

We then attempt a somewhat different approach. Instead of estimating the PDFs of the closure variables, we intend to find an approximate dynamic for the time evolution for these variables, but one that depends only on the closure variables and not on any of the components from the reduced model, therefore de-coupling the closure variables from the reduced model in one direction. We do this with a well established tool called Sparse Identification of Non-linear Dynamics (SINDy) \cite{sindy}. These tools will be discussed below.

\subsection{Variational Auto-Encoders}

Variational auto-encoders are neural networks that learn how to reconstruct data in two steps, namely encoding and decoding. The encoding process maps available data $\mathbf{x} \in \mathbb{R}^n$, into a latent vector  $ \mathbf{z}$ in a latent space $\mathbb{R}^p$, with typically $p\leq n$. The decoding process maps $\mathbf{z}$ into a reconstructed data $\mathbf{\hat{x}} \in \mathbb{R}^n$. The encoder $\mathbf{e}_\eta$ and the decoder $\mathbf{d}_\phi$, can be viewed themselves as classical neural networks with parameters (weights and biases) $\eta$ and $\phi$, but they are part of the VAE, a larger neural network with a bottleneck corresponding to the latent space, and their weights and biases are trained together. 

Variational auto-encoders differ from classical auto-encoders in the construction of the latent space. The first presents an extra imposition that $ \mathbf{z} \sim \mathcal{N}(0,I)$, i.e., that $\mathbf{z}$ is normally distributed with mean 0 and identity variance, while the latter only reconstructs the data \cite{reviewvae}. In the VAE case, the encoder produces a mean $\mu$ and a variance $\Sigma$, which are used to generate the latent variable from $\mathcal{N}(\mu,\Sigma)$, which should approximate $\mathcal{N}(0,I)$ once the network is trained.

This is a crucial element, which allows the generation of new instances of data in the same distribution as the original data by a trained VAE, simply by sampling an instance of the latent variable from $ \mathcal{N}(0,I)$ and running it through the decoder. For a VAE, we can write its loss function as 
\begin{gather}
    \mathcal{L}(\eta,\phi,\mathbf{x}) = || \mathbf{x}-\mathbf{d}_\phi(\mathbf{e}_\eta( \mathbf{x})) ||_2  + D_{KL}\big(\mathcal{N}(\mu,\Sigma)||\mathcal{N}(0,I)\big), \label{eqn:loss}
\end{gather} 
\noindent where
\begin{gather}
D_{KL}\big(\mathcal{N}(\mu,\Sigma)||\mathcal{N}(0,I)\big) =   \frac{1}{2} \sum_{i=1}^{p} \left(1 + \ln(\sigma_i^2) - \mu_i^2 - \sigma_i^2   \right), \label{eqn:divkl}
\end{gather}
\noindent for $\mu = (\mu_1, \dots ,\mu_p)^T$ and $\Sigma = \text{diag} (\sigma_1^2, \dots, \sigma_p^2)$. We are searching for the weights and biases of encoder and decoder, namely $\eta$ and $\phi$, that minimize the loss function in equation \eqref{eqn:loss}.

The expression in equation \eqref{eqn:divkl} represents the Kullback-Leibler (KL) divergence, also known as relative entropy. This metric quantifies the difference between two probability distributions. Specifically, the formula presented here applies to comparing a normal distribution with mean $\mu$ and diagonal covariance matrix $\Sigma$ to a standard normal distribution with zero mean and identity covariance matrix \cite{divkl}. Although assuming the latent variable follows a normal distribution is not strictly necessary, the availability of a closed-form expression for the KL divergence significantly simplifies the training process \cite{vae}.

\subsection{Sparse Identification for Non-linear Dynamics}

A different type of machine learning tool, SINDy relies on estimating the dynamics of a system based solely on data from trajectories. Given a trajectory $\mathbf{x}(t)$, we would like to estimate the function $\mathbf{f}(\mathbf{x},t)$ that satisfies

\begin{gather}
    \frac{d\mathbf{x}}{dt} = \mathbf{f}(\mathbf{x},t).
\end{gather}

First proposed in \cite{sindy}, the idea behind it is to give a library of functions that are computed on the trajectory points available and write each one as columns of a matrix $\Theta$, then multiply $\Theta$ by a sparse matrix of coefficients $\Xi$. This matrix of coefficients will indicate which terms from the library of functions are, in fact, present in the dynamic. We can write
\begin{gather}
    \frac{d\mathbf{x}}{dt} = \Theta \cdot \Xi. \label{eqn:dxdt}
\end{gather}

In equation  \eqref{eqn:dxdt} we can compute $\Theta$ and $\frac{d\mathbf{x}}{dt}$ numerically, so, given a sparsity parameter $c$, we can iteratively solve for $\Xi$ imposing that, at each iteration, all entries from the matrix of coefficients $\Xi$ that are below $c$ are set to zero. Equation \eqref{eqn:dxdt} can be solved by least squares \cite{sindy}, it can be regularized in many different ways \cite{conv,noisysindy,halsindy}, and it can be done by ensembles \cite{esindy}, which is the method we are applying here.

Ensemble SINDy can be implemented in several ways. One approach involves performing bootstrapping—a random sampling with replacement—on the trajectory data, on the terms of the library function, or on both. Each bootstrapped subset results in a different model, and the collection of these models forms an ensemble of SINDy models. The ensemble is typically aggregated by taking either the mean or median of the models produced. This technique is a well-established process known as bagging, a combination of "bootstrap" and "aggregating," first introduced by \cite{Breiman1996BaggingP}. 

This approach allows us to estimate inclusion probabilities for each term in the library of functions based on how often it appears on the dynamic recovered by the models in the ensemble. This can be used to quantify how well the models agree with each other, as well as defining a threshold and not include terms of the dynamic that do not appear often enough. 

The last ingredient for the SINDy approach is to turn it into a probabilistic tool, and we are going to do so by writing it as a stochastic differential equation

\begin{gather}
    d\mathbf{x} = (\Theta \cdot \Xi) dt + \sigma dW, \label{eqn:stochdxdt}
\end{gather}

\noindent where $\sigma$ is a diffusion coefficient and $W$ is a standard Brownian motion \cite{oksendal}. A similar process was conducted in \cite{Mandal2024}. This allows us to write at least part of the closure variables as a stochastic process that solves equation \eqref{eqn:stochdxdt} in the Itô sense \cite{oksendal}. We then will solve for the closure variable using a simple Euler-Maruyama scheme\cite{eulermaru}, and consider two cases: the first, where closure variables are one realization of said stochastic process, and the second, where the closure variables are a mean path of several realizations.

\section{Results} \label{sec:results}

When we choose a cutoff shell in the inertial range, the system we are trying to close does not suffer the effects of viscosity, so reduced models are of the form 
\begin{gather}
    \frac{d\mathcal{U}_N}{d\tau} = i(k_{N+1}\mathcal{U}_{N+2}\mathcal{U}_{N+1}^* -\frac{1}{2}k_{N} \mathcal{U}_{N+1}\mathcal{U}_{N-1}^* + \frac{1}{2}k_{N-1}\mathcal{U}_{N-1}\mathcal{U}_{N-2}) +\left( \xi + \xi_f \right) \mathcal{U}_N + T_m^2 k_m f_{N+m}, \nonumber \\
    N = -s,\ldots,-1. 
    \label{eqn:systemclosed}
\end{gather}

We then need to be able to compare reduced models with the fully resolved Sabra, so we will look at some statistics and observables, namely moments of order $p$, given by
\begin{gather}
    S_p(k_n) =\langle |u_n|^ p \rangle, \label{eqn:moment}
\end{gather}
\noindent for $p=2,\dots,6$, where $\langle \cdot \rangle$ is a time average, as well as energy flux through shell $n$, given by
\begin{gather}
    \Pi_n  = \operatorname{Im}(k_{n+1}u_{n+2}u_{n+1}^*u_n^* +\frac{1}{2}k_{n}u_{n+1}u_{n}^*u_{n-1}^* ). \label{eqn:fluxshell}
\end{gather}

We are also comparing PDFs of real parts, normalized by standard deviation. Despite the reduced model and closures being written for rescaled variables, all statistics are computed in terms of original variables for comparison purposes.  Multiplier phases have been initially kept at a constant value $\Delta_N = \pi/2$, which is a value that implies strict dissipation of energy at the last shell \cite{Biferale_2017}. It was included in the modeling of closure variables by the VAEs, but not by SINDy.  

The final comparison we are reporting is local slopes of $p-$order moments, given by

\begin{gather}
    \zeta_{n}^{p} = \frac{\log(S_p(k_{n+1})) - \log(S_p(k_{n}))}{\log(\lambda)}. \label{eqn:slopes}
\end{gather}

This quantity is useful to evaluate accuracy of closures based on how they compare against the same quantity $\zeta_{n,S}^{p}$ computed for the Sabra moments of order $p$. Such comparison can be commonly found in the literature \cite{sabrann,sabrann25,lstmsabra}, here we show local slopes for $p=2$, and normalized by the reference value of $\zeta_{n,S}^{p}$.

Simulations of reduced models using the VAE closures run until $\tau = 50000$, sufficient to collect robust statistics, and with a timestep $d\tau = 10^{-3}$, using a second-order Adams-Bashforth scheme. Simulations of reduced models using the SINDy closures run for the same length of time with the same timestep, but with an Euler-Maruyama scheme.

In general, to simulate equation \eqref{eqn:stochdxdt} for a given approximate dynamic and a chosen $\sigma$, a more careful choice of timestep may be required if $\sigma$ introduces intricacies such as very elaborate time and/or scale dependence. It is possible that these will have an effect on the choice of timestep, or even on the choice of numerical method. Our choice of timestep is sustained by the fact that our $\sigma$ will be chosen as constant in the next section.

\subsection{VAE}

\begin{figure}[t]
    \begin{subfigure}{.48\textwidth}
        \subcaption{Component PDF of $|\mathcal{U}_0|$}
        \includegraphics[scale=0.5]{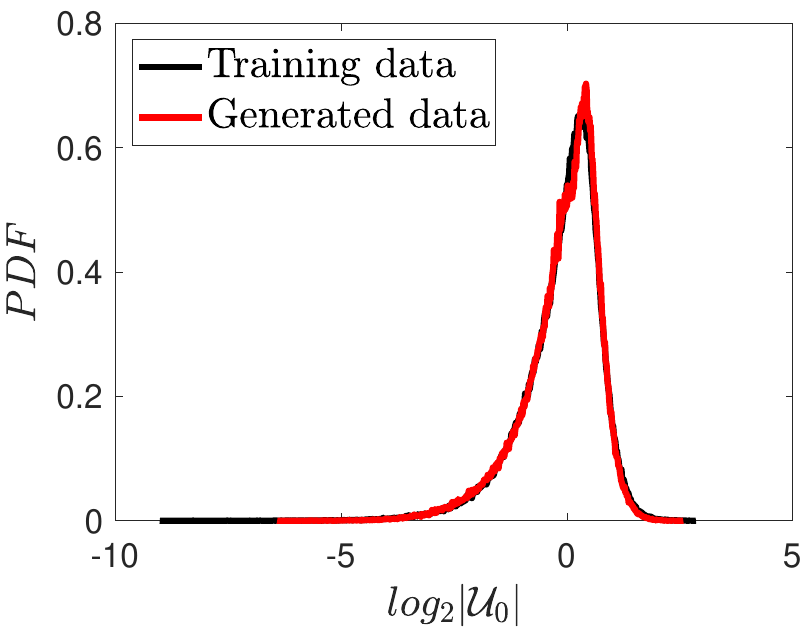}
        \label{fig:vaeU0}
    \end{subfigure}
    \begin{subfigure}{.48\textwidth}
        \subcaption{Component PDF of $|\mathcal{U}_1|$}
        \includegraphics[scale=0.5]{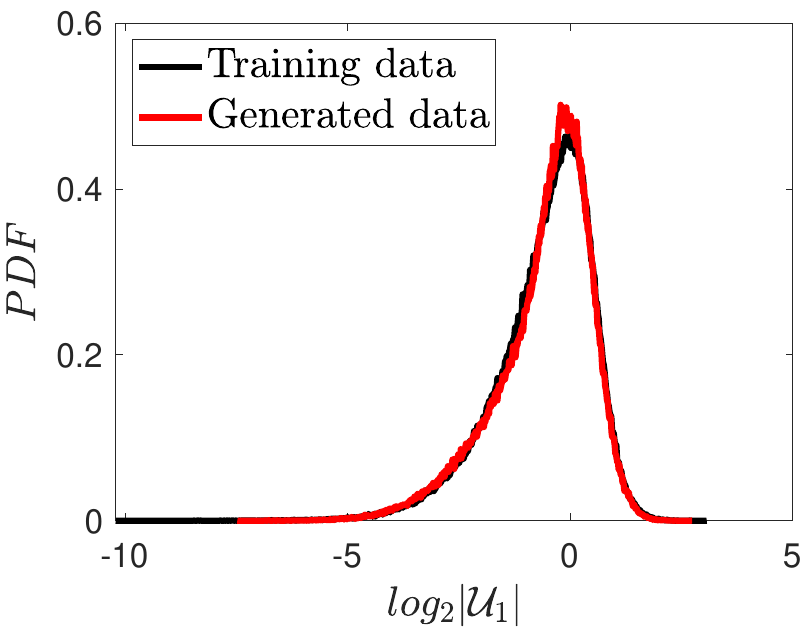}
        \label{fig:vaeU1}
    \end{subfigure}
        
    \begin{subfigure}{.48\textwidth}
        \subcaption{Joint PDF of original data}
        \includegraphics[scale=0.5]{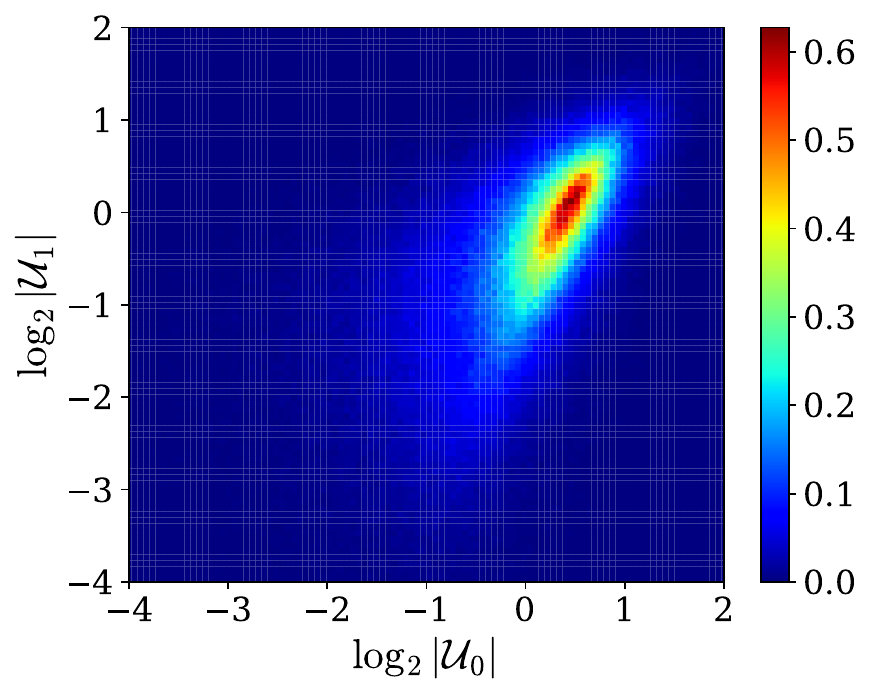}
        \label{fig:jointtrue}
    \end{subfigure}
        \begin{subfigure}{.48\textwidth}
        \subcaption{Joint PDF of data generate by the VAE-M}
        \includegraphics[scale=0.5]{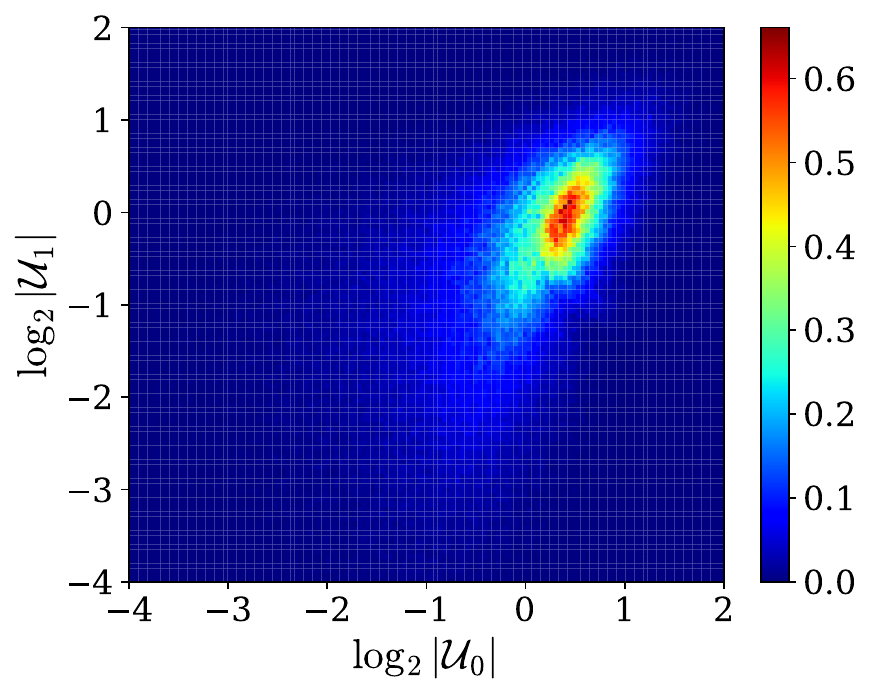}
        \label{fig:jointvae}        
    \end{subfigure}
    \caption{Comparison between original data used for training and data generated by the VAE-M.} \label{fig:vae-m}
\end{figure} 

Making use of the VAE to generate modules of the closure variables, we can write the following closure
\begin{gather} 
\mathcal{U}_{0} = 2^{y_0} e^{i(\pi/2 + \alpha_{-1} + \alpha_{-2})}, \label{eqn:closvae0}\\
\mathcal{U}_{1} = 2^{y_1} e^{i(\pi/2 + \alpha_{0} + \alpha_{-1})}, \label{eqn:closvae1}\\
(y_0,y_1) = \mathbf{d}_{\phi} (\mathbf{z}), \:\:\:\: \mathbf{z} \sim \mathcal{N}(0,\mathbf{I})
\end{gather} 

\noindent where $\alpha_N = \arg (\mathcal{U}_N)$, $\mathbf{d}_{\phi}$ is the decoder part of the trained VAE and $\mathbf{z}$ is normally distributed with mean 0 and variance identity in the same dimension  as the latent space.

The VAE trained for this problem, which will be referred from now on as VAE-M, has a total of five layers and learns data of the form $(\log_2|\mathcal{U}_{0}|,\log_2|\mathcal{U}_{1}|)$. The encoder has two layers, with 64 and 512 neurons, the latent space is two-dimensional with a layer of 128 neurons, and the decoder is a mirror of the encoder, with two layers of 512 and 64 neurons. Activation functions are all ReLU \cite{Householder1941relu} and weights and biases are initialized uniformly as per \cite{heinit}. It was trained in four sessions of $50000$ epochs with batch sizes varying from $32$ to $256$ with an Adam optimizer \cite{adam}. We can evaluate how well this density estimation is being performed by comparing original training data with data generated by the VAE-M, as can be seen in figure \ref{fig:vae-m}.

\begin{figure}[t]
    \begin{subfigure}{.24\textwidth}
        \subcaption{}
        \includegraphics[scale=0.28]{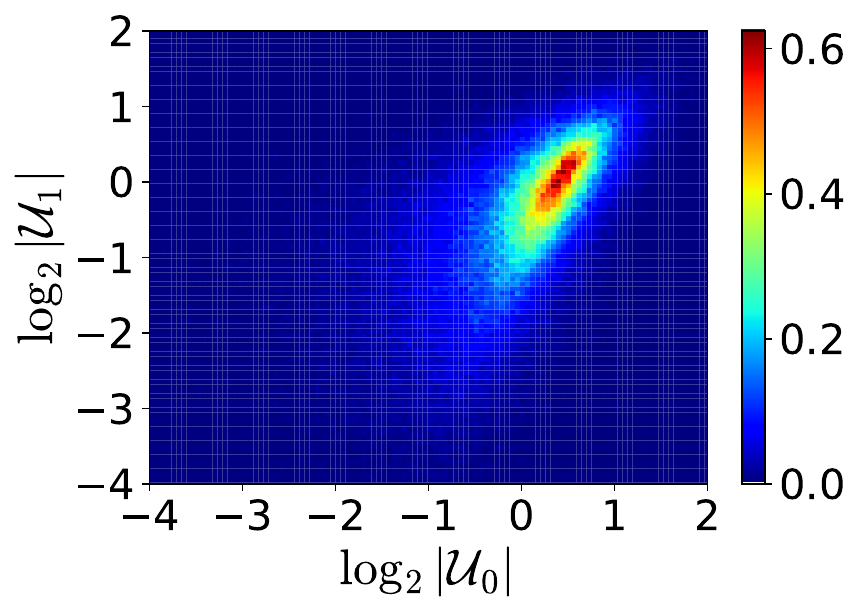}
        \label{fig:origU0U1}
    \end{subfigure}
    \begin{subfigure}{.24\textwidth}
        \subcaption{}
        \includegraphics[scale=0.28]{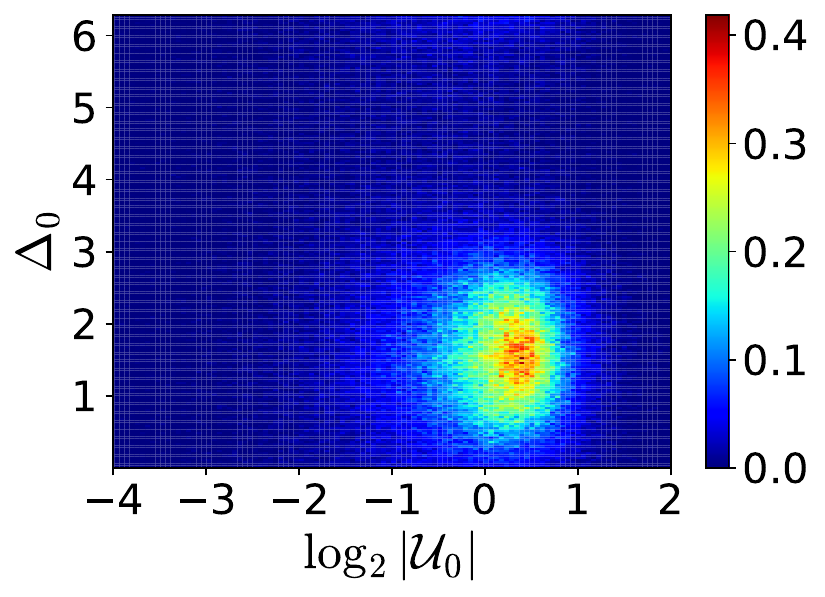}
        \label{fig:origU0d0}
    \end{subfigure}
    \begin{subfigure}{.24\textwidth}
        \subcaption{}
        \includegraphics[scale=0.28]{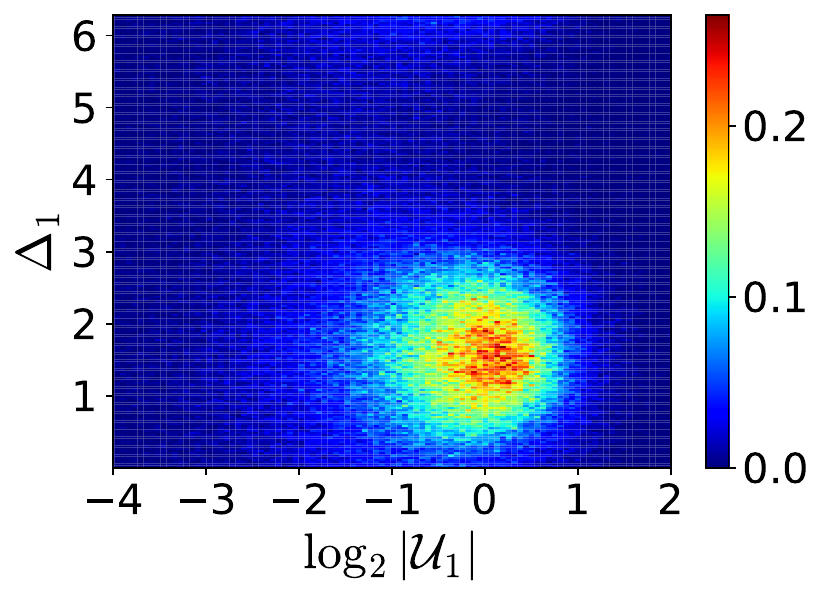}
        \label{fig:origU1d1}
    \end{subfigure}
    \begin{subfigure}{.24\textwidth}
        \subcaption{}
        \includegraphics[scale=0.28]{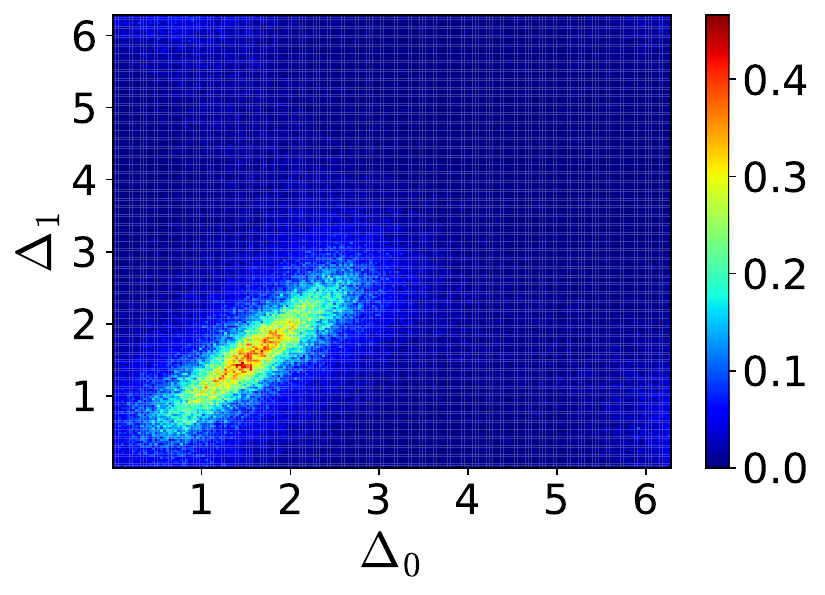}
        \label{fig:origd0d1}
    \end{subfigure}
        
    \begin{subfigure}{.24\textwidth}
        \subcaption{}
        \includegraphics[scale=0.28]{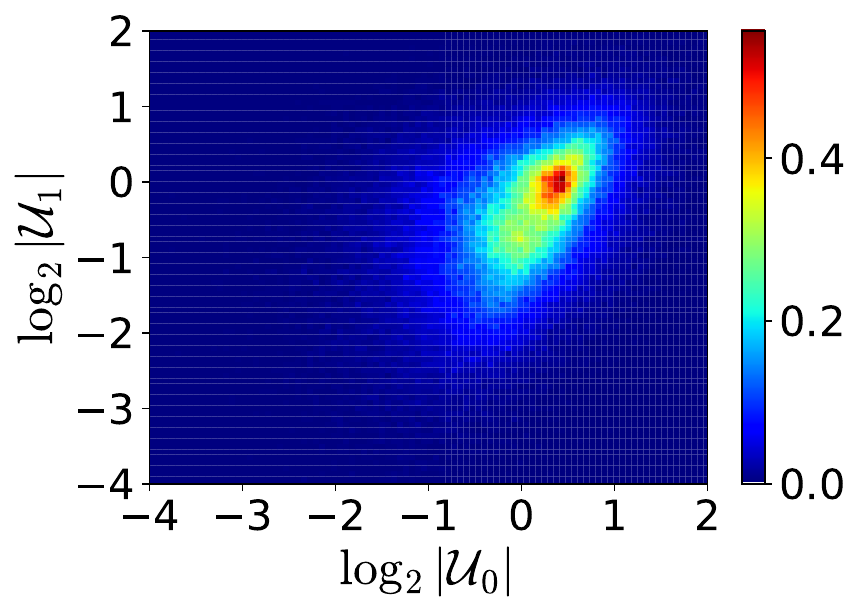}
        \label{fig:vaeU0U1}
    \end{subfigure}
        \begin{subfigure}{.24\textwidth}
        \subcaption{}
        \includegraphics[scale=0.28]{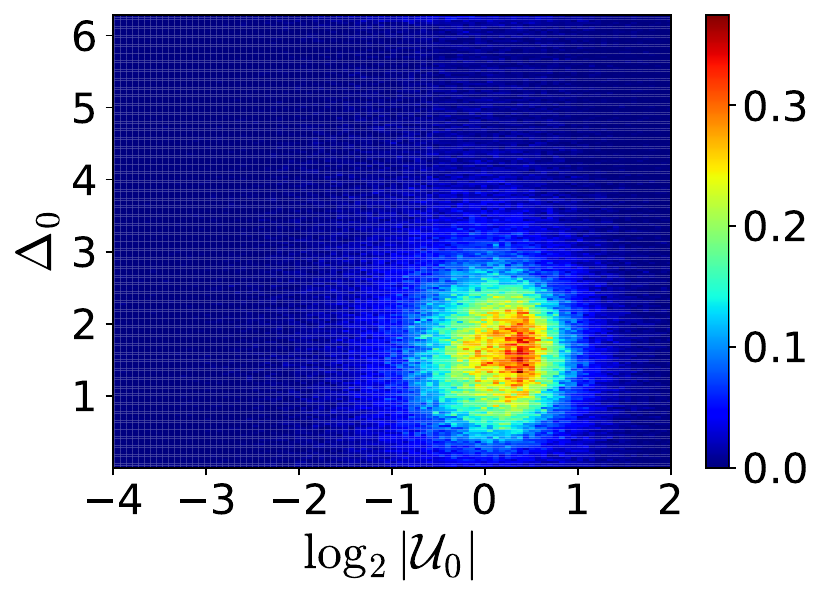}
        \label{fig:vaeU0d0}        
    \end{subfigure}
    \begin{subfigure}{.24\textwidth}
        \subcaption{}
        \includegraphics[scale=0.28]{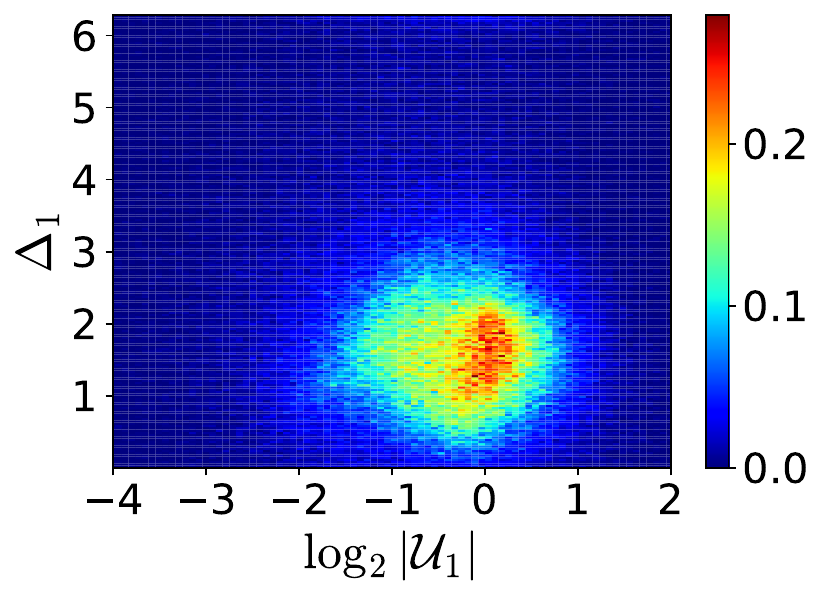}
        \label{fig:vaeU1d1}
    \end{subfigure}
    \begin{subfigure}{.24\textwidth}
        \subcaption{}
        \includegraphics[scale=0.28]{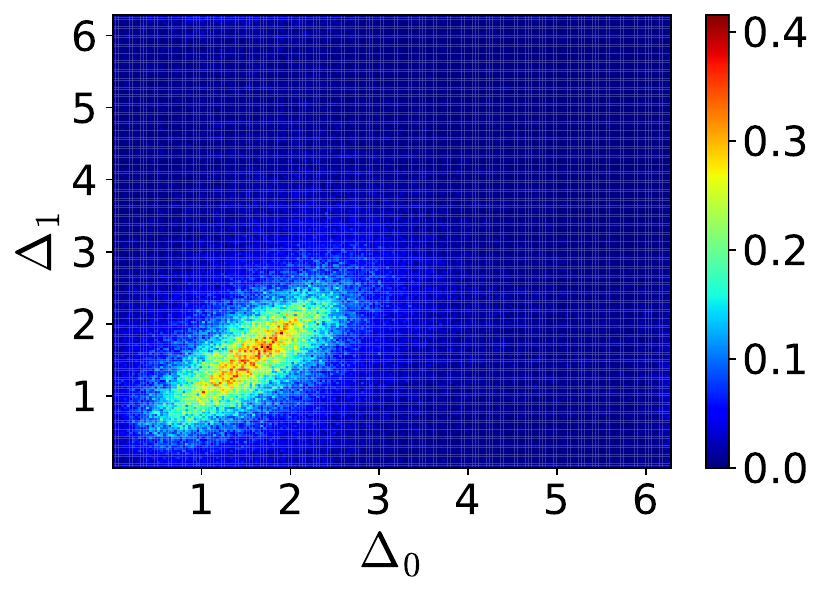}
        \label{fig:vaed0d1}
    \end{subfigure}
    \caption{Upper row \ref{fig:origU0U1}---\ref{fig:origd0d1} presents the pairwise densities of the original data, bottom row \ref{fig:vaeU0U1}---\ref{fig:vaed0d1} presents the same pairwise densities for data generated by the VAE-P.} \label{fig:vae-p}
\end{figure}

Since multiplier's phases effectively control how energy dissipates \cite{Biferale_2017}, we know that, without including phases in our modelling of the closure variables, we can not correctly estimate the closure variables. We then write a closure that includes phases, of the form
\begin{gather} 
\mathcal{U}_{0} = 2^{y_0} e^{i(y_1 + \alpha_{-1} + \alpha_{-2})}, \label{eqn:closvae_withphases0}\\
\mathcal{U}_{1} = 2^{y_2} e^{i(y_3 + \alpha_{0} + \alpha_{-1})}, \label{eqn:closvae_withphases1}\\
(y_0,y_1,y_2,y_3) = \bar{\mathbf{d}}_{\phi} (\mathbf{z}), \:\:\:\: \mathbf{z} \sim \mathcal{N}(0,\mathbf{I}),
\end{gather} 

\noindent where $\bar{\mathbf{d}}_{\phi}$ is the decoder part of a different VAE, from now on referred to as VAE-P, trained on data of the form \\ $(\log_2|\mathcal{U}_{0}|,\log_2|\mathcal{U}_{1}|,\Delta_0,\Delta_1)$. The VAE-P encoder has two layers of 1024 neurons, the latent space has dimension six with layer of 128 neurons and a decoder of the same size as the encoder. Weights and biases are still uniformly initialized, but the activation function is now a sine function, which was chosen to accommodate for the periodic nature of the phase densities, seen in \ref{fig:delta01}. It was trained in 12 sessions of $50000$ epochs with batch sizes varying from $32$ to $256$ with an Adam optimizer. A comparison between data generated by the VAE-P and the original data can be seen in figure \ref{fig:vae-p}.

\begin{figure}[t!]
\begin{subfigure}{.333\textwidth}
  \centering
  \subcaption{$n=12$}
  \includegraphics[scale=0.42]{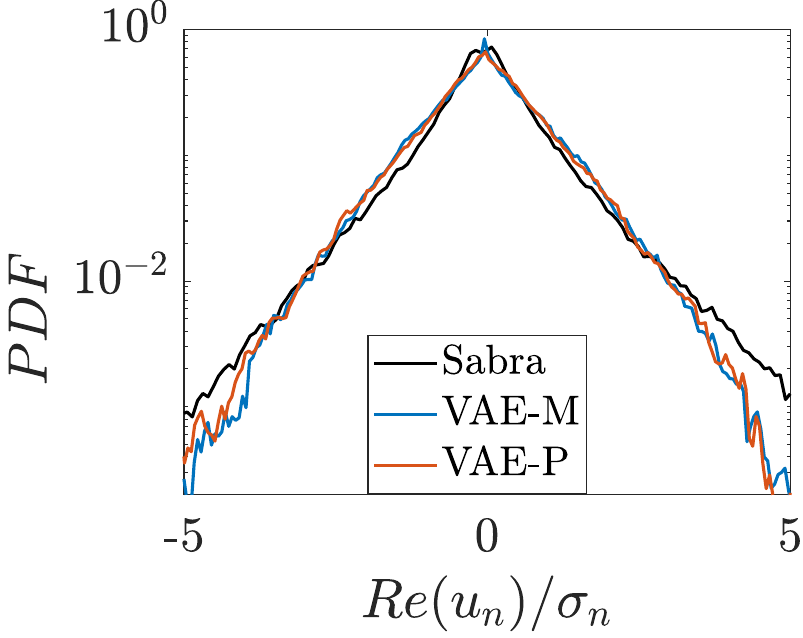}
  \label{fig:re12}
\end{subfigure}%
\begin{subfigure}{.333\textwidth}
  \centering
  \subcaption{$n=13$}
  \includegraphics[scale=0.42]{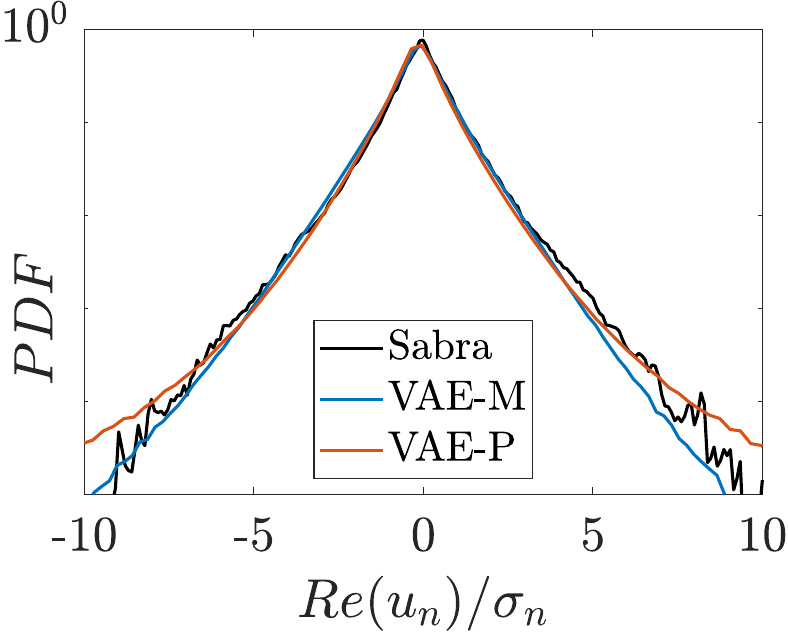}
  \label{fig:re13}
\end{subfigure}%
\begin{subfigure}{.333\textwidth}
  \centering
  \subcaption{$n=14$}
  \includegraphics[scale=0.42]{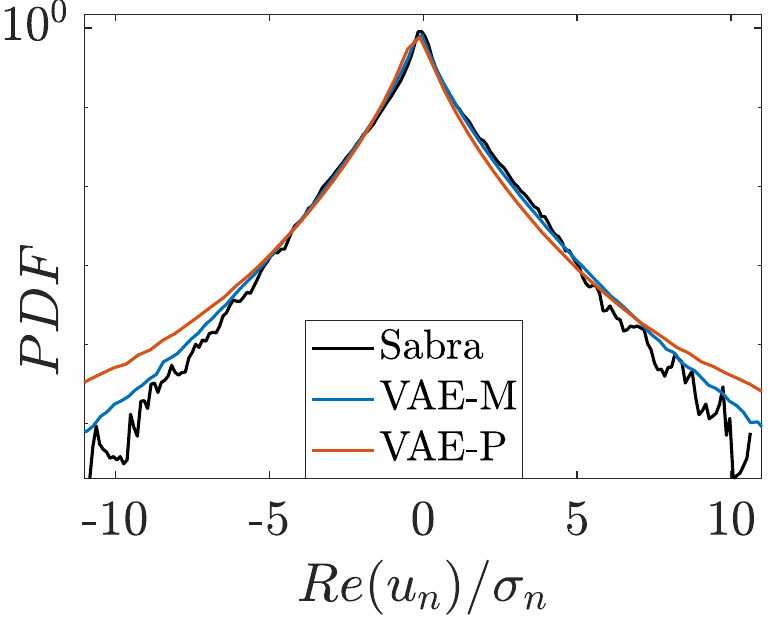}
  \label{fig:re14}
\end{subfigure}%

\begin{subfigure}{.5\textwidth}
  \centering
  \subcaption{VAE-M}
  \includegraphics[scale=0.60]{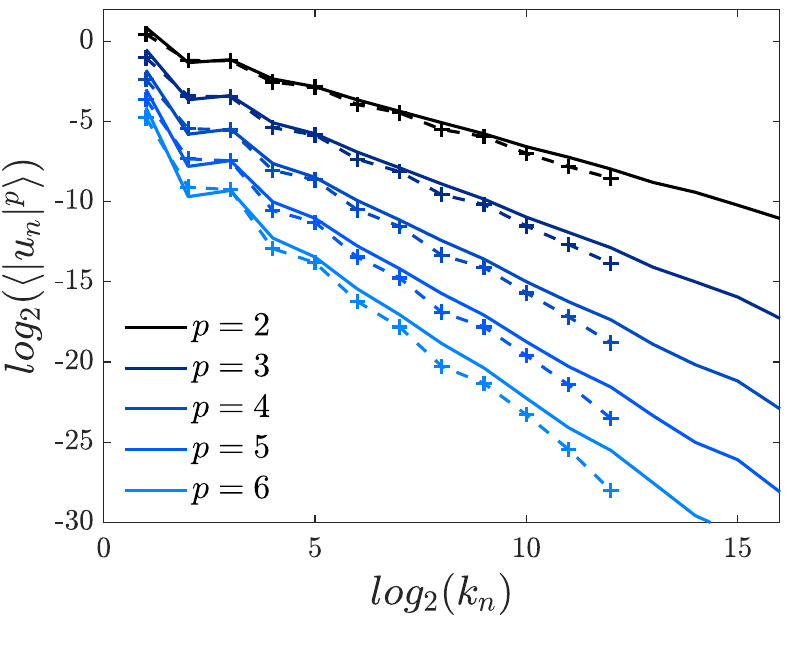}
  \label{fig:mjoint}
\end{subfigure}%
\begin{subfigure}{.5\textwidth}
  \centering
  \subcaption{VAE-P}
  \includegraphics[scale=0.60]{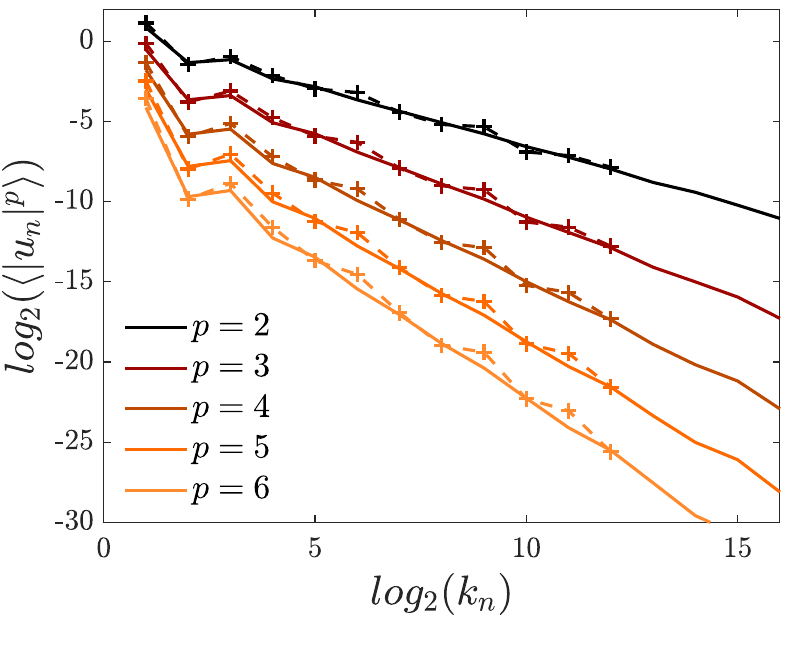}
  \label{fig:pjoint}
\end{subfigure}%

\begin{subfigure}{.333\textwidth}
  \centering
  \subcaption{$n=10$}
  \includegraphics[scale=0.42]{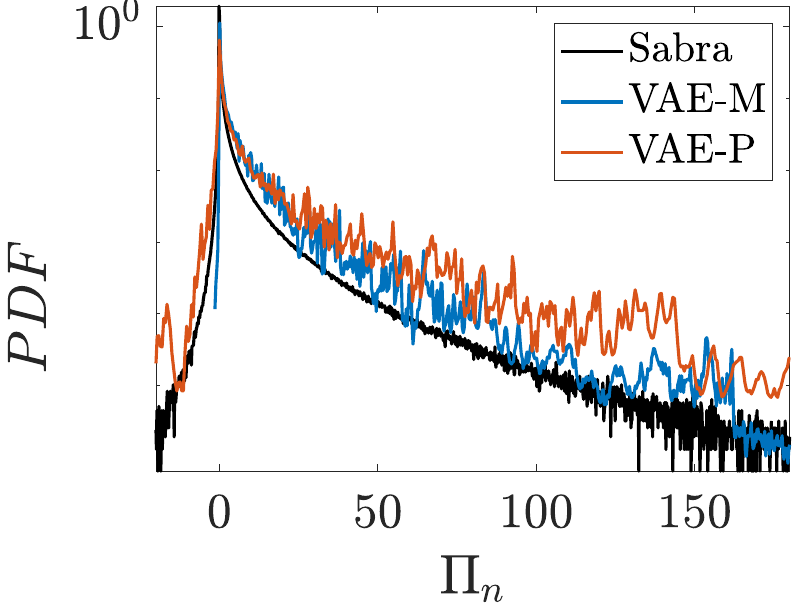}
  \label{fig:flx9}
\end{subfigure}%
\begin{subfigure}{.333\textwidth}
  \centering
  \subcaption{$n=11$}
  \includegraphics[scale=0.42]{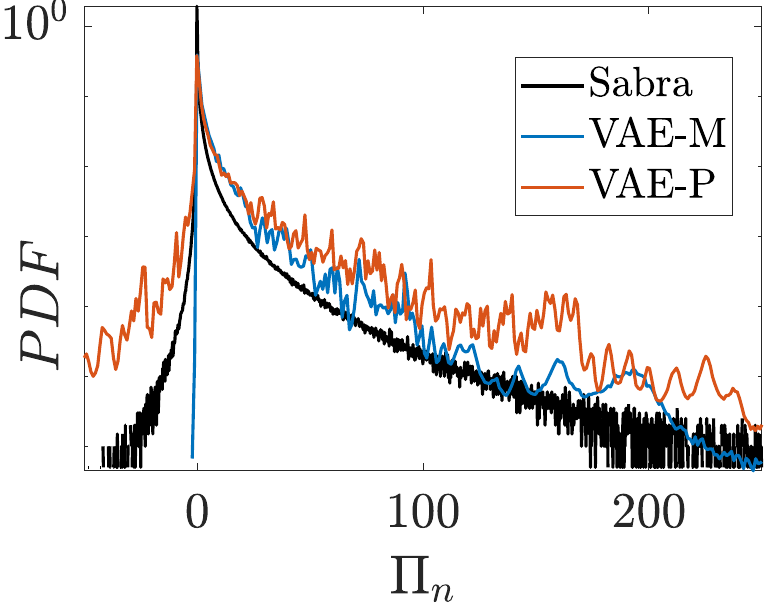}
  \label{fig:flx11}
\end{subfigure}%
\begin{subfigure}{.333\textwidth}
  \centering
  \subcaption{$n=12$}
  \includegraphics[scale=0.42]{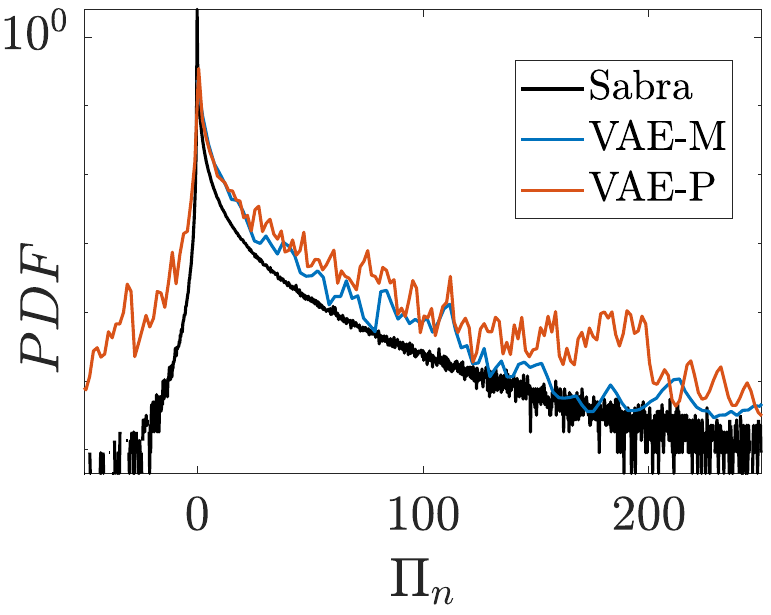}
  \label{fig:flx12}
\end{subfigure}%
\caption{Cutoff is at $s=12$. Performance of closures based on trained VAE-M, which only includes modules, and VAE-P, which also includes phases. Comparison is made based on Sabra statistics of the fully resolved model. Figures \ref{fig:re12}---\ref{fig:re14} show normalized PDFs of real parts for the last shell of the reduced model and for the closure variables. Figures \ref{fig:flx9}---\ref{fig:flx12} show energy flux PDFs for the last shells of the reduced model. Figures \ref{fig:mjoint} and \ref{fig:pjoint} display moments of orders 2 up to 6 for reduced model simulations performed using VAE-M and VAE-P respectively. Both are compared to Sabra moments in solid lines.}
\label{fig:vae12}
\end{figure}

\begin{figure}[t!]
\begin{subfigure}{.333\textwidth}
  \centering
  \subcaption{$n=9$}
  \includegraphics[scale=0.42]{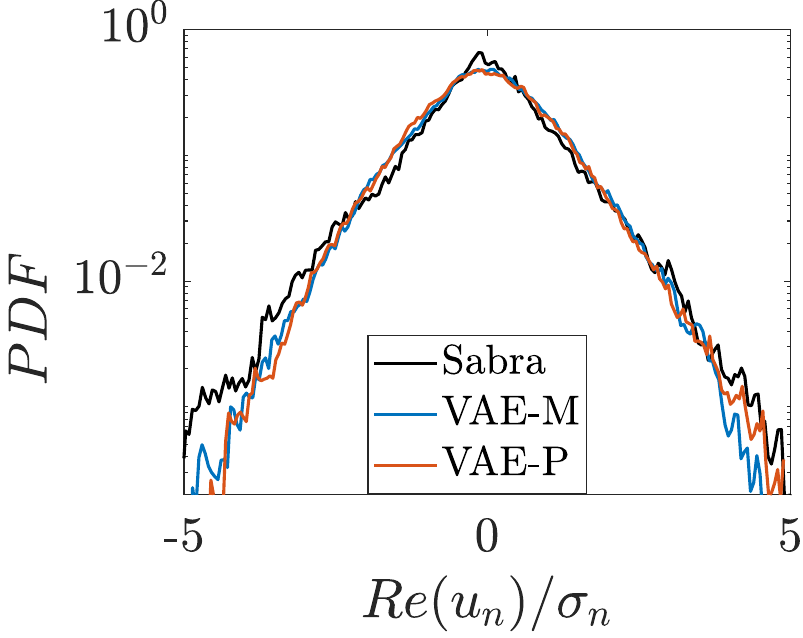}
  \label{fig:vae9re9}
\end{subfigure}%
\begin{subfigure}{.333\textwidth}
  \centering
  \subcaption{$n=10$}
  \includegraphics[scale=0.42]{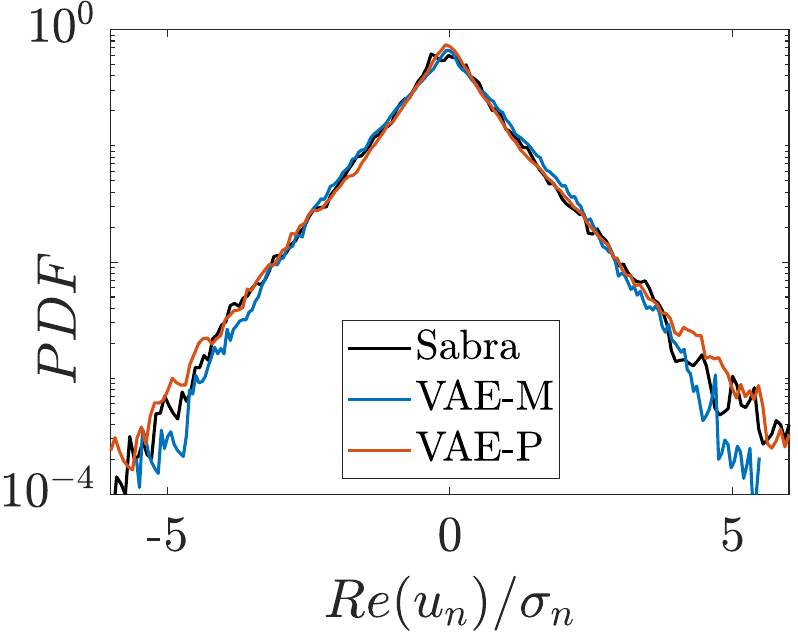}
  \label{fig:vae9re10}
\end{subfigure}%
\begin{subfigure}{.333\textwidth}
  \centering
  \subcaption{$n=11$}
  \includegraphics[scale=0.42]{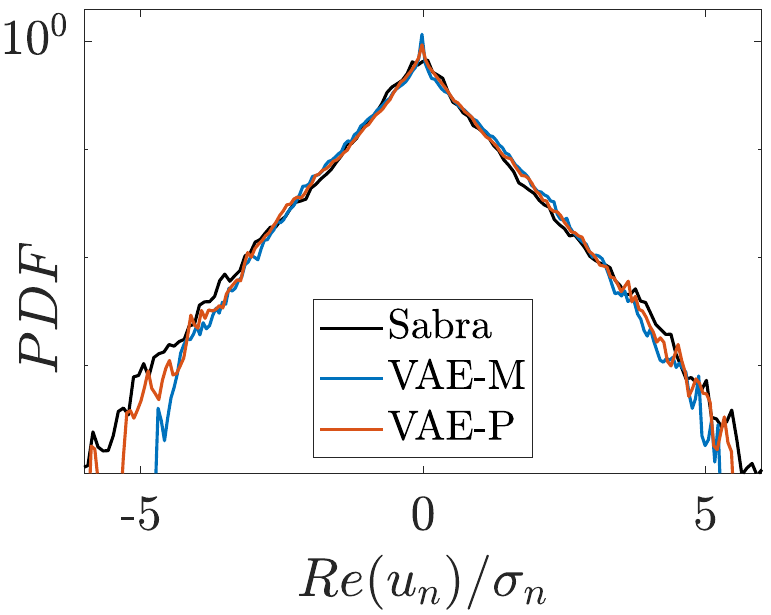}
  \label{fig:vae9re11}
\end{subfigure}%

\begin{subfigure}{.5\textwidth}
  \centering
  \subcaption{VAE-M}
  \includegraphics[scale=0.60]{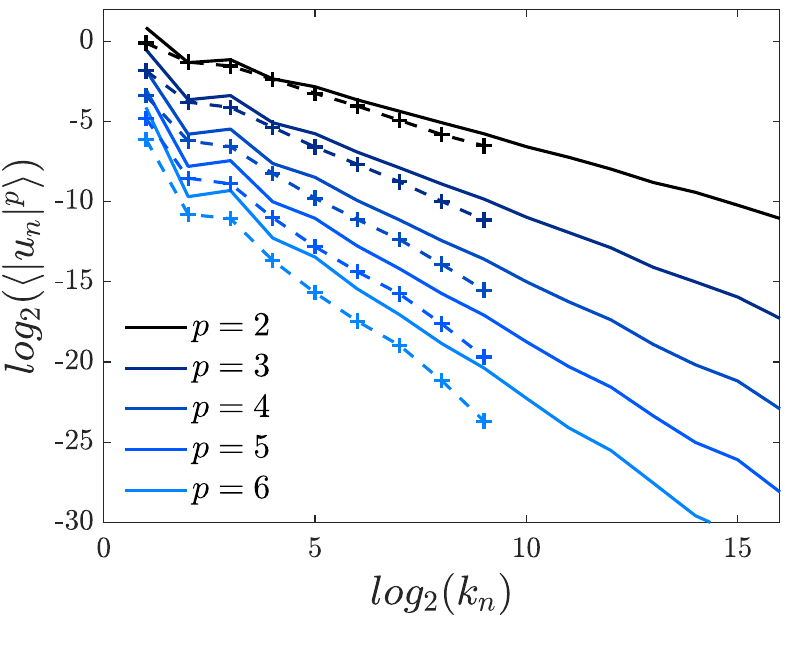}
  \label{fig:vae9mjoint9}
\end{subfigure}%
\begin{subfigure}{.5\textwidth}
  \centering
  \subcaption{VAE-P}
  \includegraphics[scale=0.60]{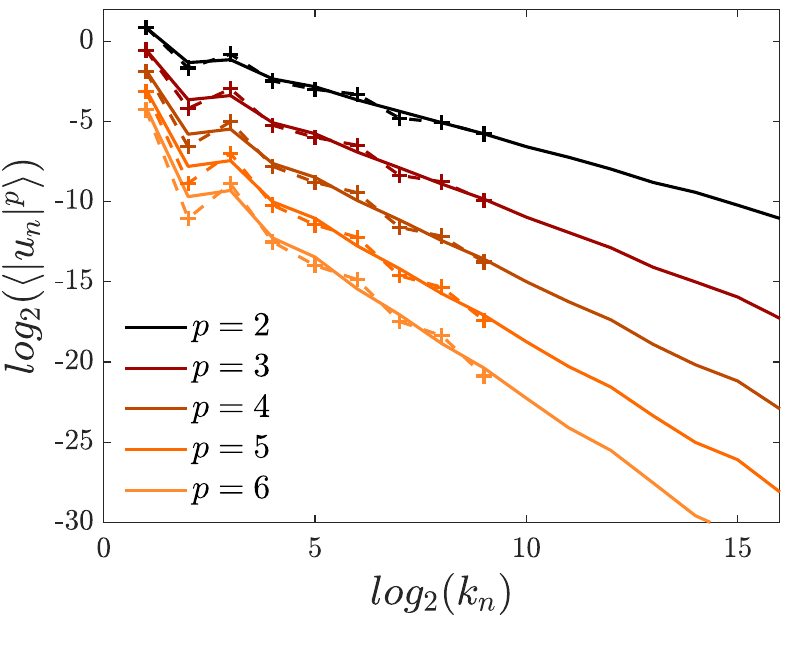}
  \label{fig:vae9pjoint9}
\end{subfigure}%

\begin{subfigure}{.333\textwidth}
  \centering
  \subcaption{$n=7$}
  \includegraphics[scale=0.35]{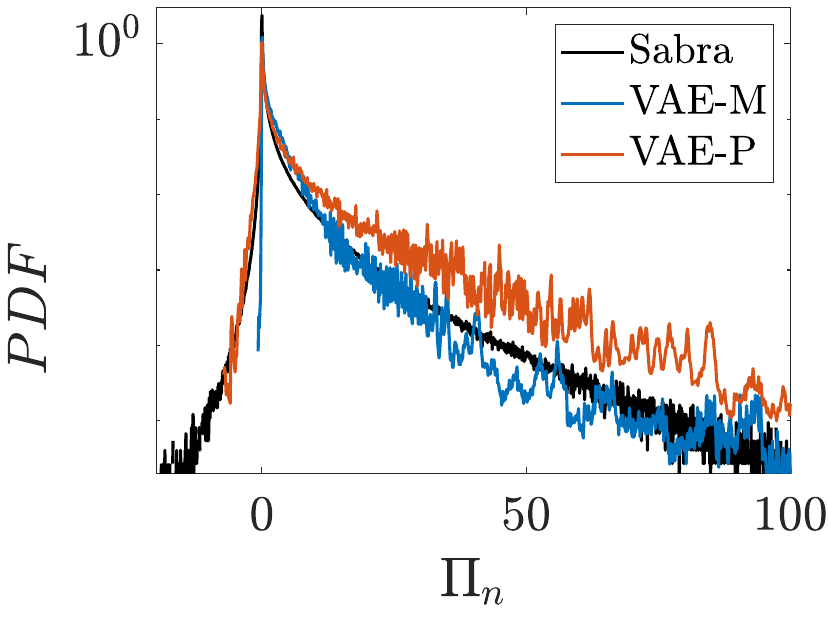}
  \label{fig:vae9flx7}
\end{subfigure}%
\begin{subfigure}{.333\textwidth}
  \centering
  \subcaption{$n=8$}
  \includegraphics[scale=0.34]{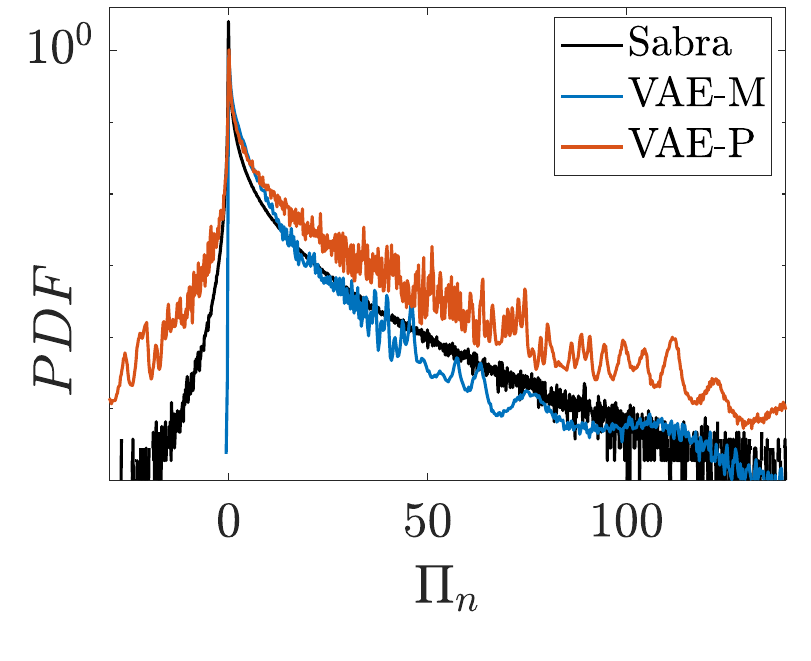}
  \label{fig:vae9flx8}
\end{subfigure}%
\begin{subfigure}{.333\textwidth}
  \centering
  \subcaption{$n=9$}
  \includegraphics[scale=0.35]{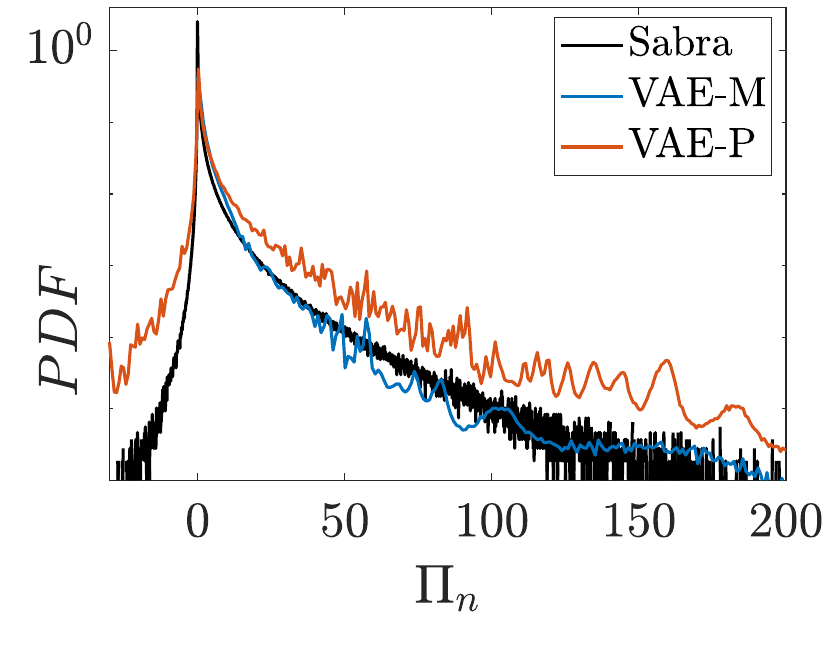}
  \label{fig:vae9flx9}
\end{subfigure}%
\caption{Cutoff is at $s=9$. Performance of closures based on trained VAE-M, which only includes modules, and VAE-P, which also includes phases. Comparison is made based on Sabra statistics of the fully resolved model. Figures \ref{fig:vae9re9}---\ref{fig:vae9re9} show normalized PDFs of real parts for the last shell of the reduced model and for the closure variables. Figures \ref{fig:vae9flx7}---\ref{fig:vae9flx9} show energy flux PDFs for the last shells of the reduced model. Figures \ref{fig:vae9mjoint9} and \ref{fig:vae9pjoint9} display moments of orders 2 up to 6 for reduced model simulations performed using VAE-M and VAE-P respectively. Both are compared to Sabra moments in solid lines.}
\label{fig:vae9}
\end{figure}

In these two cases, we are using the VAE-M and the VAE-P to generate the closure variables at each time step in a simulation of the reduced model from equation \eqref{eqn:systemclosed}, although pre-generating the data for the closure variables will yield the same results. We are also using the same VAE-M and VAE-P to generate closure variables for a different cut-off without retraining the tools. The performance of these closures can be seen in the comparison of statistics in figures \ref{fig:vae12} and \ref{fig:vae9}.

In figures \ref{fig:re12}---\ref{fig:re14} we see show normalized PDFs of real parts for the cutoff shell $s=12$ and the subsequent closure variables. In figures \ref{fig:flx9}---\ref{fig:flx12} we see energy flux PDFs for the last shells of the reduced model, computed according to equation \eqref{eqn:fluxshell}. In figures \ref{fig:mjoint} and \ref{fig:pjoint} we see moments of orders 2 up to 6, computed according to equation \eqref{eqn:moment}. Note that, as scales progress in the inertial range towards the dissipative regime, Sabra PDFs of real parts become more and more heavy tailed \cite{lucashell}, which is also recovered by our model. In energy flux PDFs we can see a clear difference between closures that include phase modeling and closures that only model modules. If no phase modeling is present on a closure, meaning, if multipliers phases are fixed on their most probable (strictly dissipating) value of $\Delta_0=\Delta_1 = \pi/2$, it severely underestimates the back-scattering energy flux. This can be clearly seen in \ref{fig:flx12} by the complete absence of a left leg at the PDF. On the other hand, our closure that includes phases seems to always overcompensate on this matter.

Now, with a change in cutoff, we see in figures \ref{fig:vae9re9}---\ref{fig:vae9re11} normalized PDFs of real parts for the cutoff shell $s=9$ and the subsequent closure variables. In figures \ref{fig:vae9flx7}---\ref{fig:vae9flx9} we see energy flux PDFs for the last shells of the reduced model as well as moments of orders 2 up to 6 in figures \ref{fig:vae9mjoint9} and \ref{fig:vae9pjoint9}. It is crucial to note that the change of cutoff was done solely in the simulation of the reduced model. Both VAEs were not retrained and are exactly the same networks that generated the closure variables in the case of cutoff $s=12$. Firstly, we can see that the change of cutoffs does not impact qualitatively the behavior of these closures. We also see, more pronouncedly in figures \ref{fig:mjoint} and \ref{fig:pjoint}, that not including phases in the modeling also underestimates moments, with a slight difference in slope. On the other hand, including phases in the modeling leads to a more accurate estimation of moments, albeit a slightly more oscillating one.

\subsection{SINDy}

In this approach the idea is to de-couple the closure variables to evolve them independently of the reduced system, and we do that based on data from the fully resolved system. In general, it is true that
\begin{gather} 
\frac{d\mathcal{U}_{0}}{d\tau} = e^{i \alpha_0 } \frac{d|\mathcal{U}_0|}{d\tau} + i\mathcal{U}_0 \frac{d \alpha_0}{d\tau}, \label{eqn:dudtau0}\\
\frac{d\mathcal{U}_{1}}{d\tau} = e^{i \alpha_1 } \frac{d|\mathcal{U}_1|}{d\tau} + i\mathcal{U}_1 \frac{d \alpha_1}{d\tau}. \label{eqn:dudtau1}
\end{gather} 

\noindent provided $|\mathcal{U}_N|$ stays smooth for $N=0,1$. By keeping multipliers phases at $\Delta_0=\Delta_1 = \pi/2$, from equation \eqref{eqn:deltaa}, we have
\begin{gather}
    \alpha_N = \frac{\pi}{2} + \alpha_{N-1} + \alpha_{N-2},\nonumber \label{eqn:dist2}
\end{gather}

\noindent so the second term on the right-hand side of \cref{eqn:dudtau0,eqn:dudtau1} can be computed from the pre-existing data, while the first ones, containing derivatives of moduli, are to be estimated via SINDy.

\begin{figure}[t!]
\begin{subfigure}{.333\textwidth}
  \centering
  \subcaption{}
  \includegraphics[scale=0.41]{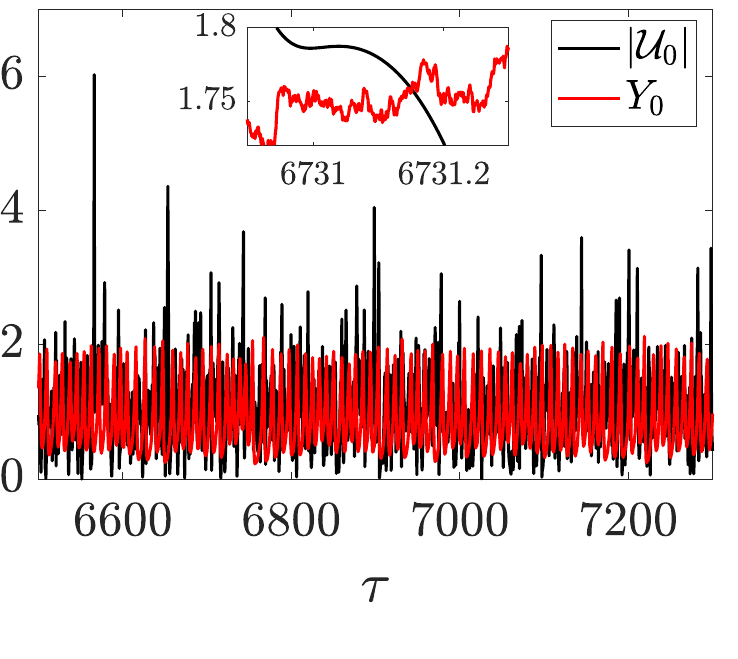}
  \label{fig:sindyu0}
\end{subfigure}%
\begin{subfigure}{.333\textwidth}
  \centering
  \subcaption{}
  \includegraphics[scale=0.41]{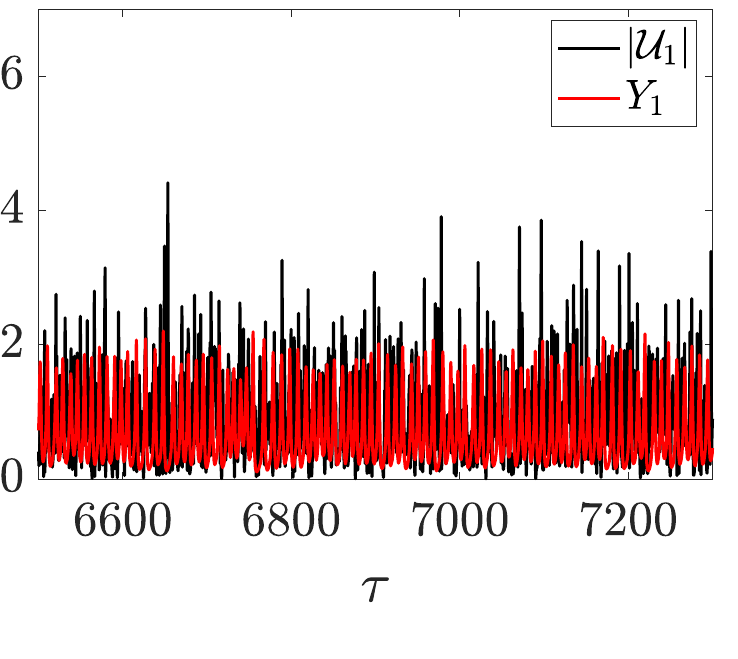}
  \label{fig:sindyu1}
\end{subfigure}%
\begin{subfigure}{.333\textwidth}
  \centering
  \subcaption{}
  \includegraphics[scale=0.41]{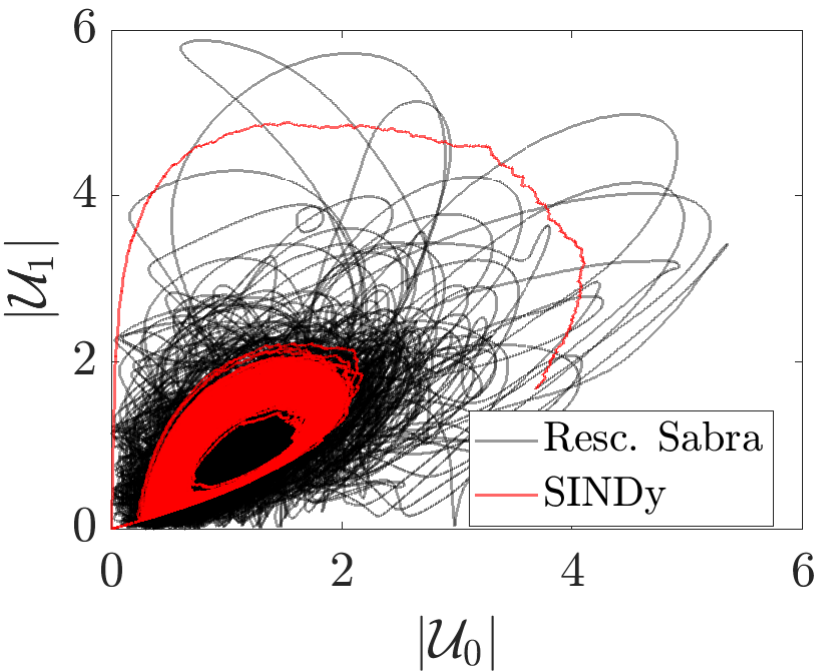}
  \label{fig:portrait}
\end{subfigure}%

\begin{subfigure}{.5\textwidth}
  \centering
  \subcaption{}
  \includegraphics[scale=0.64]{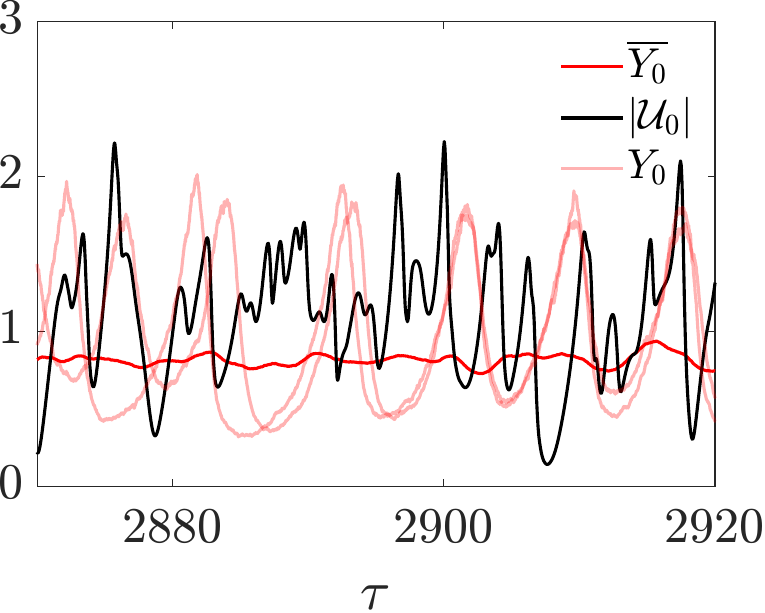}
  \label{fig:mean0}
\end{subfigure}%
\begin{subfigure}{.5\textwidth}
  \centering
  \subcaption{}
  \includegraphics[scale=0.64]{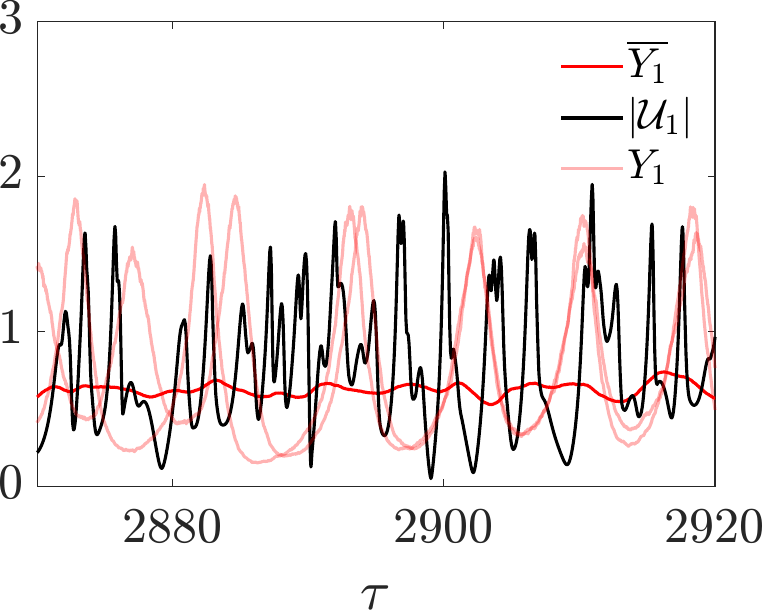}
  \label{fig:mean1}
\end{subfigure}
\caption{Top row shows one realization of the stochastic process that solves \cref{eqn:stochsindy0,eqn:stochsindy1} in red, with a comparison between the rescaled Sabra variables in black. Time evolution, with an inset for detail, can be seen in figures \cref{fig:sindyu0,fig:sindyu1}, while figure \ref{fig:portrait} shows a phase portrait of long time behavior until $\tau = 20000$. Figures \ref{fig:mean0} and \ref{fig:mean1} in the bottom row show two realization of the same process in pale red, with the mean path described in \cref{eqn:meansindy0,eqn:meansindy1} displayed in bold red (ensemble average over 100 realizations), against the black rescaled Sabra variables, until $\tau = 5000$.} \label{fig:eye}
\end{figure}

There are two considerations to be made here. The first one is that modulus functions do not necessarily remain smooth, so it is slightly unclear how to approach estimating a derivative that may or may not be discontinuous via, say, polynomial functions. This is a valid concern, but our data, after suffering the rescaling we proposed, indeed does not show any obvious points of non-differentiability. The minimum in either trajectories of $|\mathcal{U}_0|$ and $|\mathcal{U}_1|$ is larger than $0.002$ and the finite differences approximation to the derivative does not present any jump-like discontinuities. Examples of trajectories of modulus can be seen in figures \ref{fig:mean0} and \ref{fig:mean1}.

The second aspect to be considered is the fact that Sabra presents chaos \cite{Biferale_2017} and one can not see strange attractors in any dynamical systems with less than 3 dimensions as a consequence of the Poincaré-Bendixson theorem \cite{poinc}, the closest we can find is a limit cycle. 

For a sparsity parameter of 0.3, a Ridge regularization term with a parameter of 1, data bootstrapping to create an ensemble of 60 SINDy models (averaging coefficients with an inclusion probability of 0.6), and a function library with polynomials up to degree 5, SINDy identified the following dynamics:

\begin{gather}
    \frac{dy_0}{d\tau} =  -1.26y_0y_1 - y_0^2 -0.21y_0y_1^2 -0.46y_0^2y_1 -0.24y_0^3 \label{eqn:detsindy0} \\
    \frac{dy_1}{d\tau} = 0.66y_0- 1.77y_1 -0.78y_0y_1. \label{eqn:detsindy1} 
\end{gather}


\noindent for $y_0(\tau) = |\mathcal{U}_0|$ and $y_1(\tau) = |\mathcal{U}_1|$. The solution for \cref{eqn:detsindy0,eqn:detsindy1} is deterministic. So, for $\sigma = 0.05$ we write

\begin{gather}
    dY_0 =  \big( -1.26Y_0Y_1 - Y_0^2 -0.21Y_0Y_1^2 -0.46Y_0^2Y_1 -0.24Y_0^3 \big) d\tau  + \sigma dW_{0}\label{eqn:stochsindy0} \\
    dY_1 = \big(0.66Y_0- 1.77Y_1 -0.78Y_0Y_1\big) + \sigma dW_{1} \label{eqn:stochsindy1} 
\end{gather}

\noindent where $W_0$ and $W_1$ are standard one-dimensional Brownian Motion. From here we develop two closure models. The first is written as 

\begin{gather} 
\mathcal{U}_{0} = Y_0 e^{i(\pi/2 + \alpha_{-1} + \alpha_{-2})}, \label{eqn:sindy0}\\
\mathcal{U}_{1} = Y_1 e^{i(\pi/2 + \alpha_{0} + \alpha_{-1})}, \label{eqn:sindy1}
\end{gather} 

\noindent where $Y_0$ and $Y_1$ are one realization of the stochastic process that solves the stochastic differential equations \cref{eqn:stochsindy0,eqn:stochsindy1}. This is now a probabilist model that was not trained to reproduce the statistics presented in figure \ref{fig:absu0u1}, but instead attempts to evolve our closure variables in an approximate dynamics. We will refer to this closure as One-Path.

\begin{figure}[t!]
\begin{subfigure}{.333\textwidth}
  \centering
  \subcaption{$n=12$}
  \includegraphics[scale=0.42]{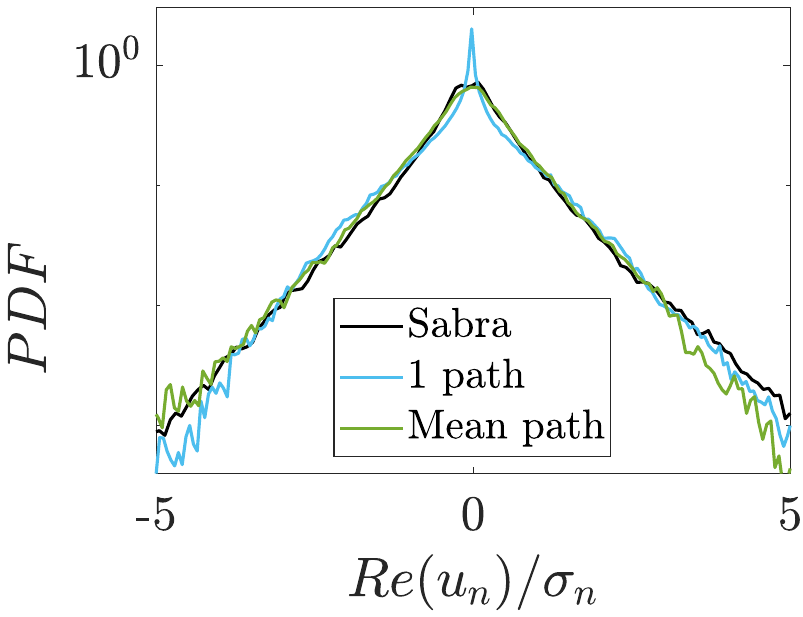}
  \label{fig:sindyre12}
\end{subfigure}%
\begin{subfigure}{.333\textwidth}
  \centering
  \subcaption{$n=13$}
  \includegraphics[scale=0.42]{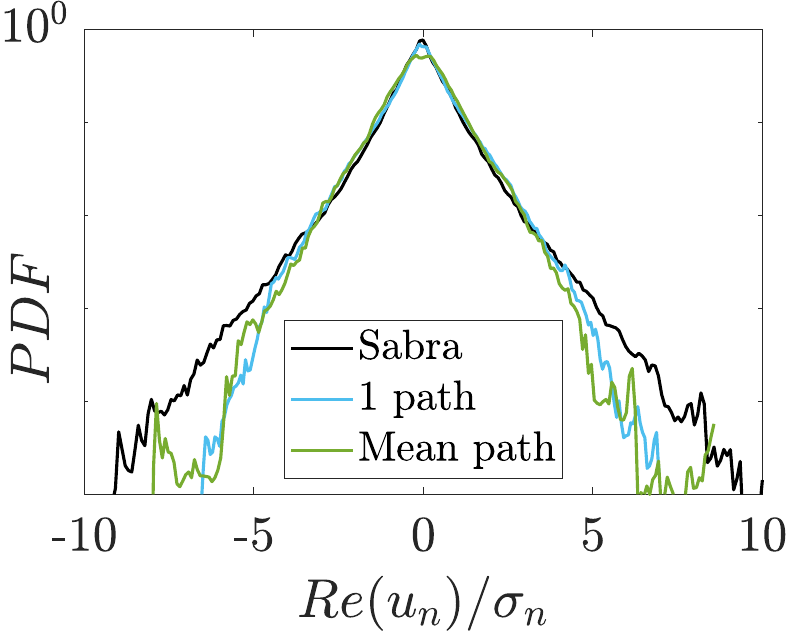}
  \label{fig:sindyre13}
\end{subfigure}%
\begin{subfigure}{.333\textwidth}
  \centering
  \subcaption{$n=14$}
  \includegraphics[scale=0.42]{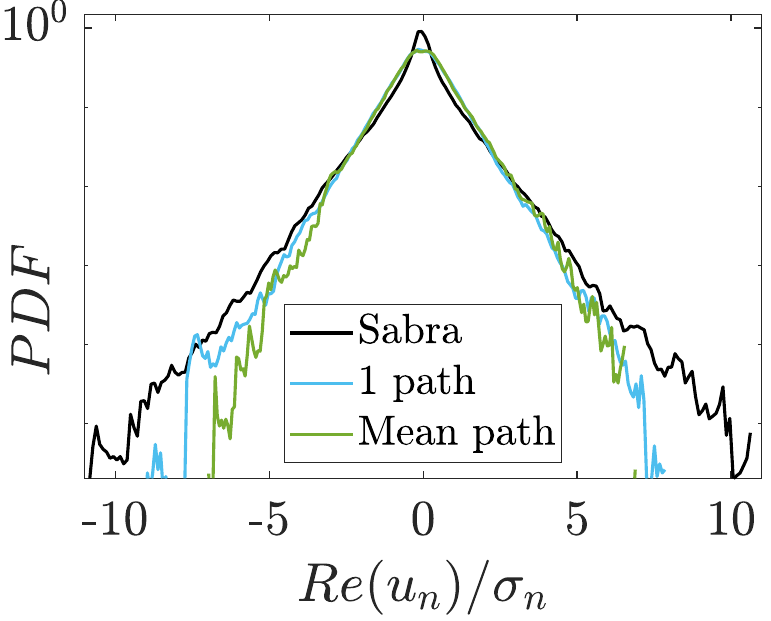}
  \label{fig:sindyre14}
\end{subfigure}%

\begin{subfigure}{.5\textwidth}
  \centering
  \subcaption{One Path}
  \includegraphics[scale=0.55]{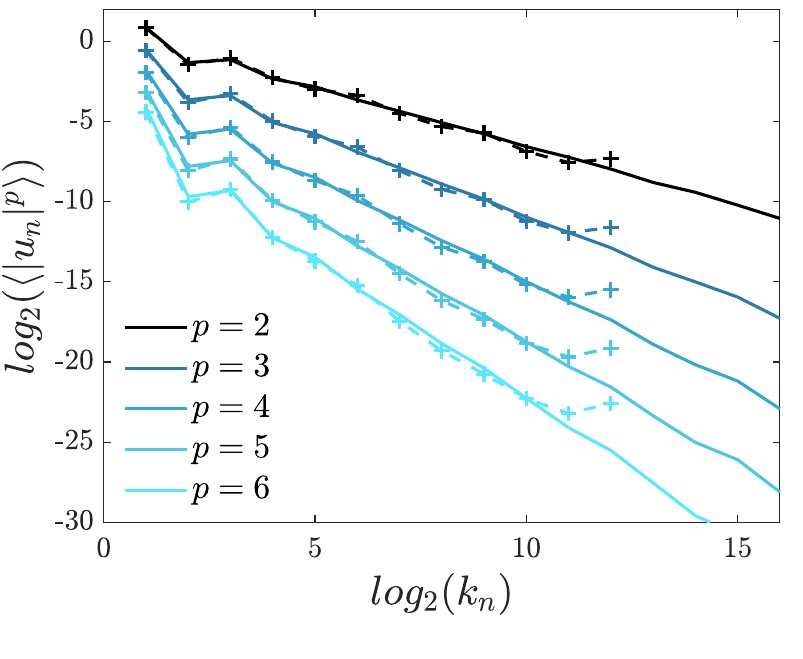}
  \label{fig:sindymjoint}
\end{subfigure}%
\begin{subfigure}{.5\textwidth}
  \centering
  \subcaption{Mean path}
  \includegraphics[scale=0.57]{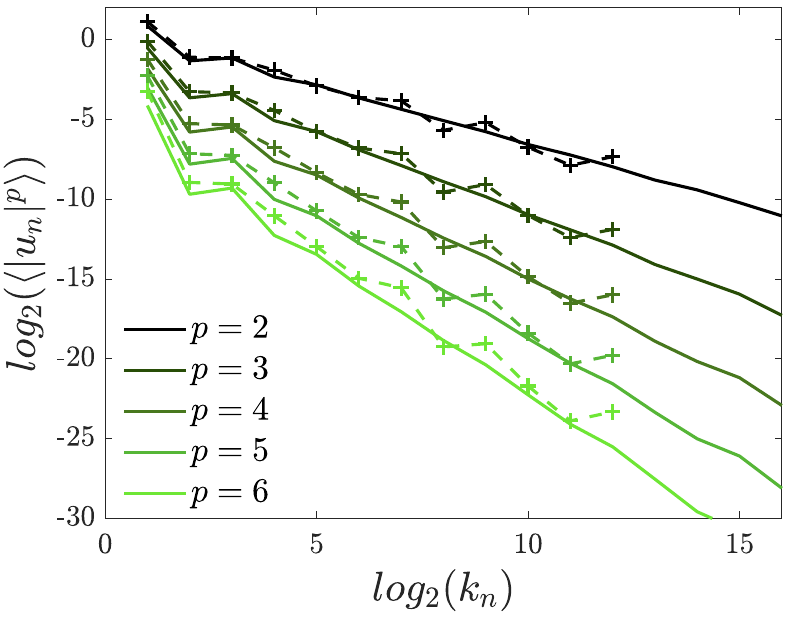}
  \label{fig:sindypjoint}
\end{subfigure}%

\begin{subfigure}{.333\textwidth}
  \centering
  \subcaption{$n=10$}
  \includegraphics[scale=0.36]{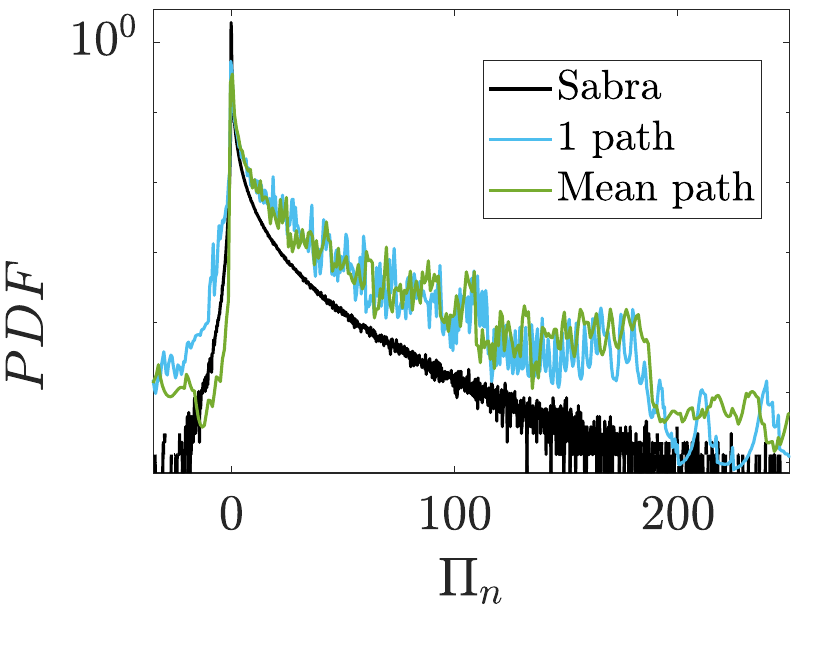}
  \label{fig:sindyflx10}
\end{subfigure}%
\begin{subfigure}{.333\textwidth}
  \centering
  \subcaption{$n=11$}
  \includegraphics[scale=0.37]{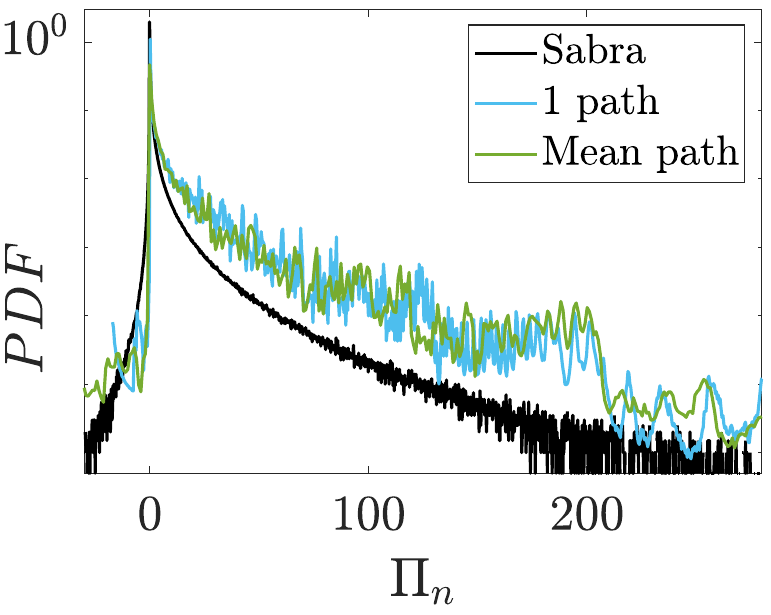}
  \label{fig:sindyflx11}
\end{subfigure}%
\begin{subfigure}{.333\textwidth}
  \centering
  \subcaption{$n=12$}
  \includegraphics[scale=0.37]{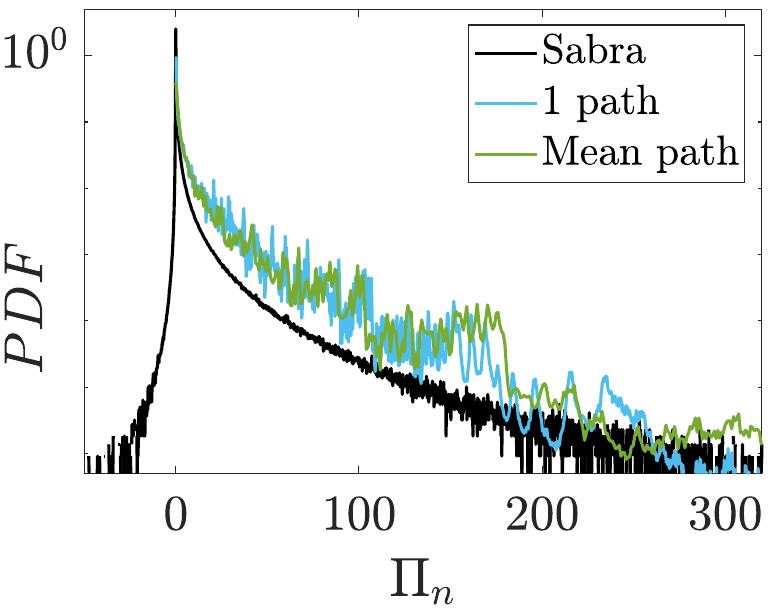}
  \label{fig:sindyflx12}
\end{subfigure}%
\caption{Cut-off shell is $s=12$. Performance of closures based on the stochastic version of the approximate dynamics found by SINDy, both using a single realization and taking a mean path. Comparison is made based on Sabra statistics of the fully resolved model. Figures \ref{fig:sindyre12}---\ref{fig:sindyre14} show normalized PDFs of real parts for the last shell of the reduced model and for the closure variables. Figures \ref{fig:sindyflx10}---\ref{fig:sindyflx12} show energy flux PDFs for the last shells of the reduced model. Figures \ref{fig:sindymjoint} and \ref{fig:sindypjoint} display moments of orders 2 up to 6 for reduced model simulations performed using one realization and a mean of several realizations, respectively. Both are compared to Sabra moments in solid lines.}
\label{fig:sindys}
\end{figure}

The second closure we will write as

\begin{gather} 
\mathcal{U}_{0} = \overline{Y_0} e^{i(\pi/2 + \alpha_{-1} + \alpha_{-2})}, \label{eqn:meansindy0}\\
\mathcal{U}_{1} = \overline{Y_1} e^{i(\pi/2 + \alpha_{0} + \alpha_{-1})}, \label{eqn:meansindy1}
\end{gather} 

\noindent where $\overline{Y_0} = \langle Y_0^j \rangle_*$ and $\overline{Y_1} = \langle Y_1^j \rangle_*$ with $j=1,\cdots,10$. Here, $\langle \cdot \rangle_*$ is an ensemble average and each pair $\big(Y_0^j,Y_1^j\big)$ is one realization of the stochastic process that solves \cref{eqn:stochsindy0,eqn:stochsindy1}, all evolved from the same initial condition until the same final time. This gives us a type of mean path, which is then used as moduli of closure variables to evolve our reduced models in time. This closure is referred to here as Mean-Path. Numerical solutions to the system described in \cref{eqn:stochsindy0,eqn:stochsindy1} can be seen in figure \ref{fig:eye}, while the performance of these closures can be seen in figures \ref{fig:sindys} and \ref{fig:sindys}.

\begin{figure}[t!]
\begin{subfigure}{.333\textwidth}
  \centering
  \subcaption{$n=9$}
  \includegraphics[scale=0.41]{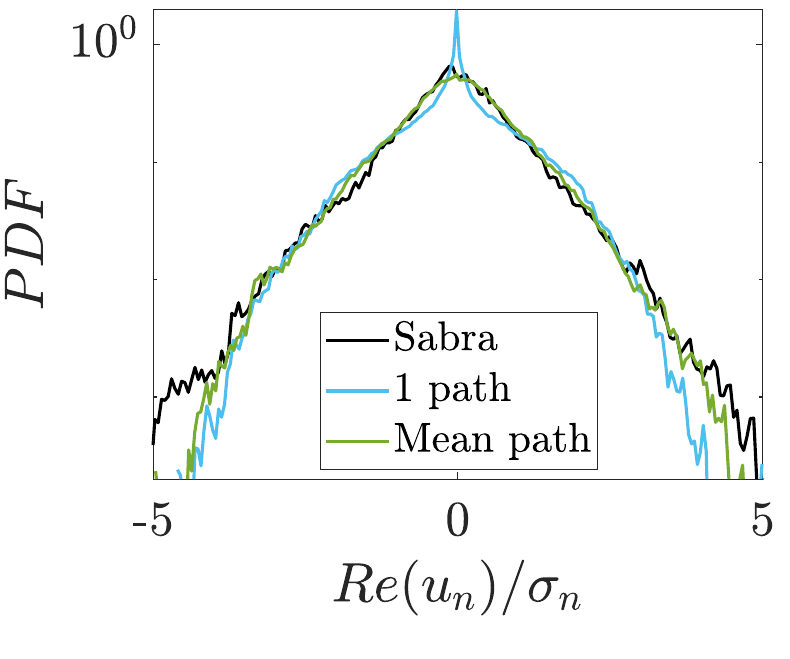}
  \label{fig:sindyre9}
\end{subfigure}%
\begin{subfigure}{.333\textwidth}
  \centering
  \subcaption{$n=10$}
  \includegraphics[scale=0.41]{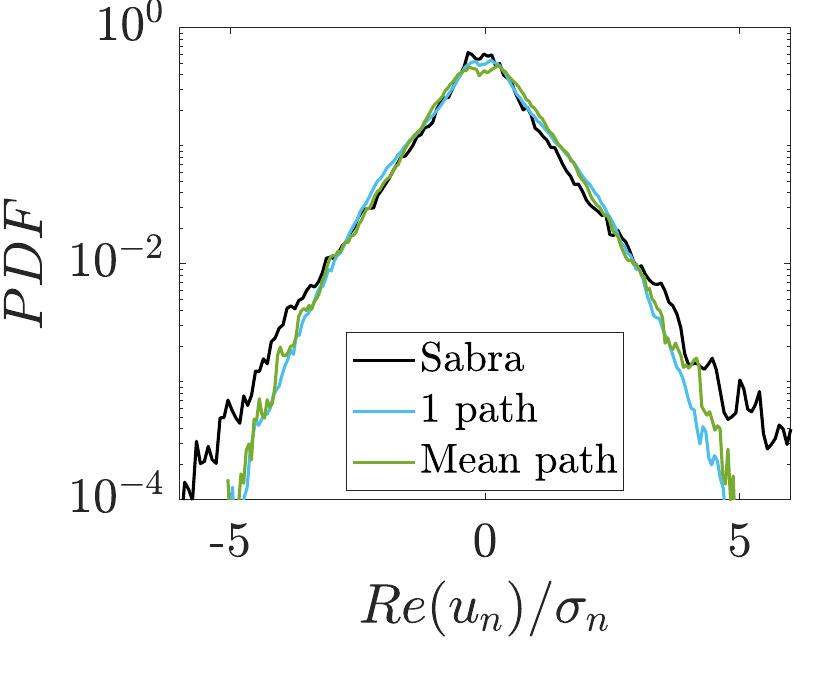}
  \label{fig:sindyre10}
\end{subfigure}%
\begin{subfigure}{.333\textwidth}
  \centering
  \subcaption{$n=11$}
  \includegraphics[scale=0.41]{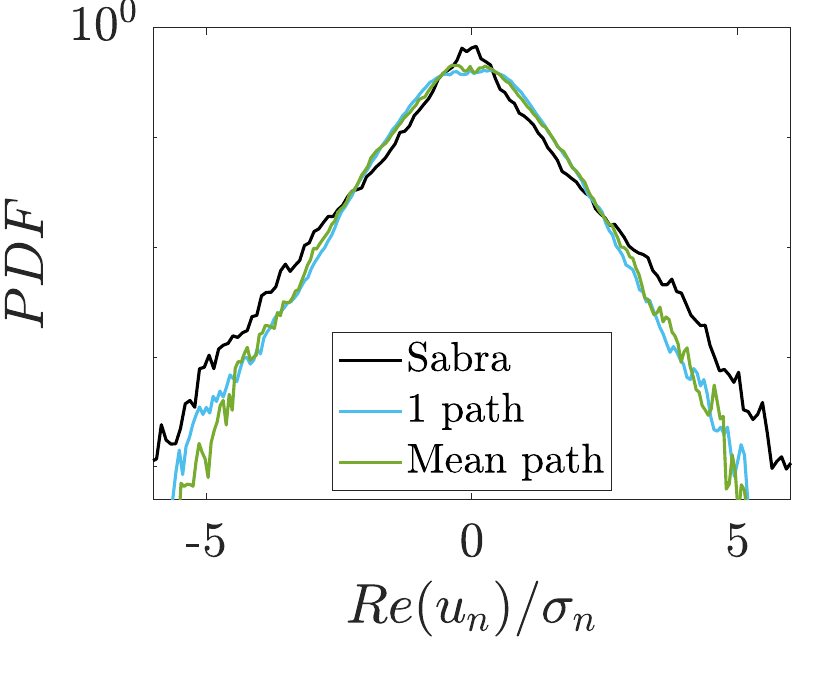}
  \label{fig:sindyre11}
\end{subfigure}%

\begin{subfigure}{.5\textwidth}
  \centering
  \subcaption{One Path}
  \includegraphics[scale=0.6]{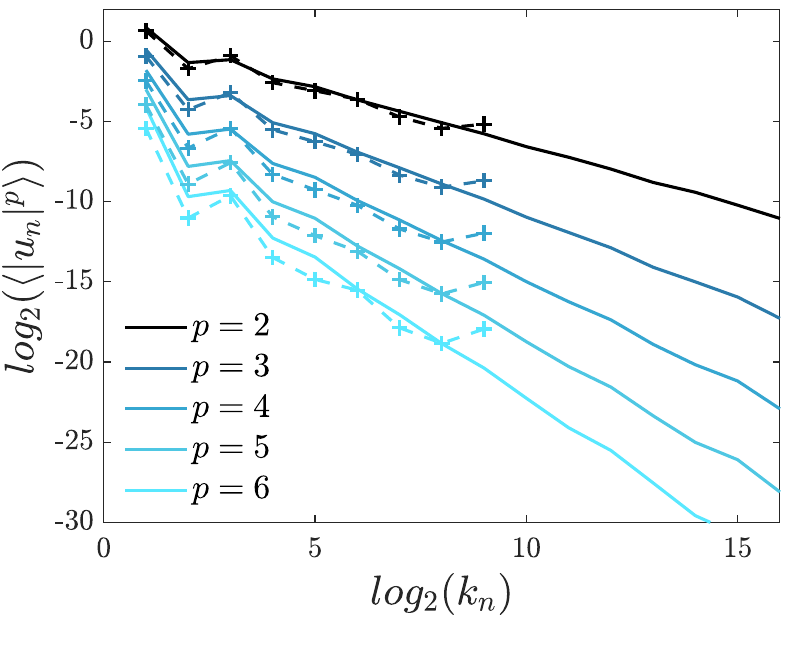}
  \label{fig:sindymjoint9}
\end{subfigure}%
\begin{subfigure}{.5\textwidth}
  \centering
  \subcaption{Mean path}
  \includegraphics[scale=0.6]{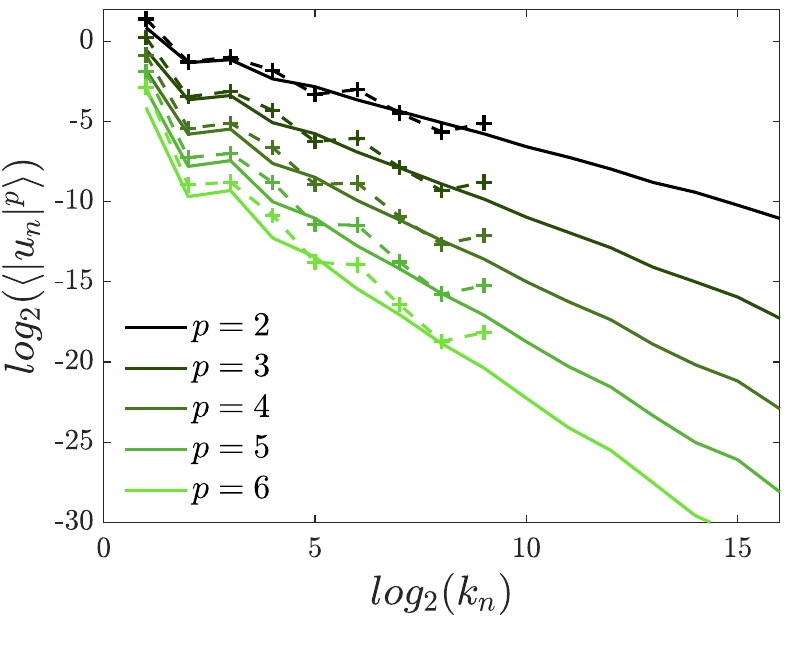}
  \label{fig:sindypjoint9}
\end{subfigure}%

\begin{subfigure}{.333\textwidth}
  \centering
  \subcaption{$n=7$}
  \includegraphics[scale=0.41]{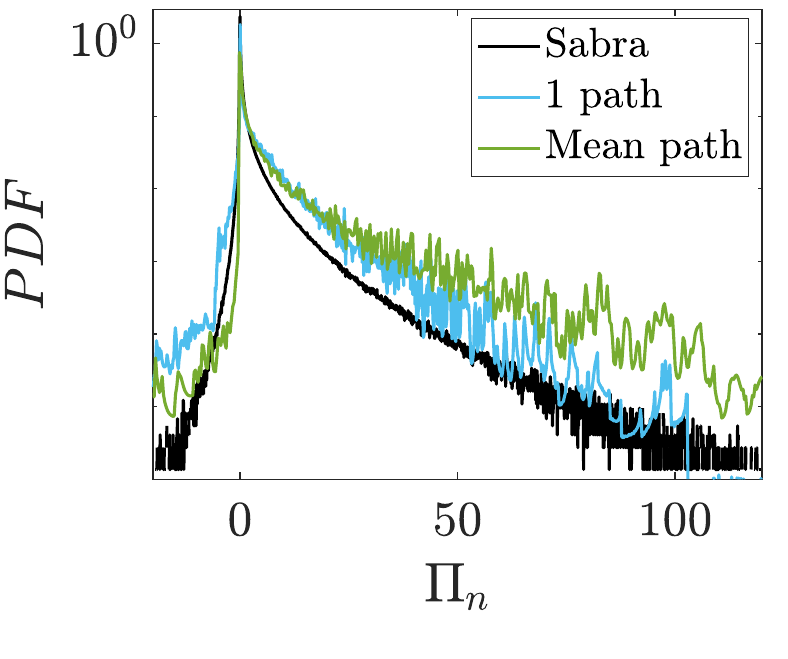}
  \label{fig:sindyflx7}
\end{subfigure}%
\begin{subfigure}{.333\textwidth}
  \centering
  \subcaption{$n=8$}
  \includegraphics[scale=0.41]{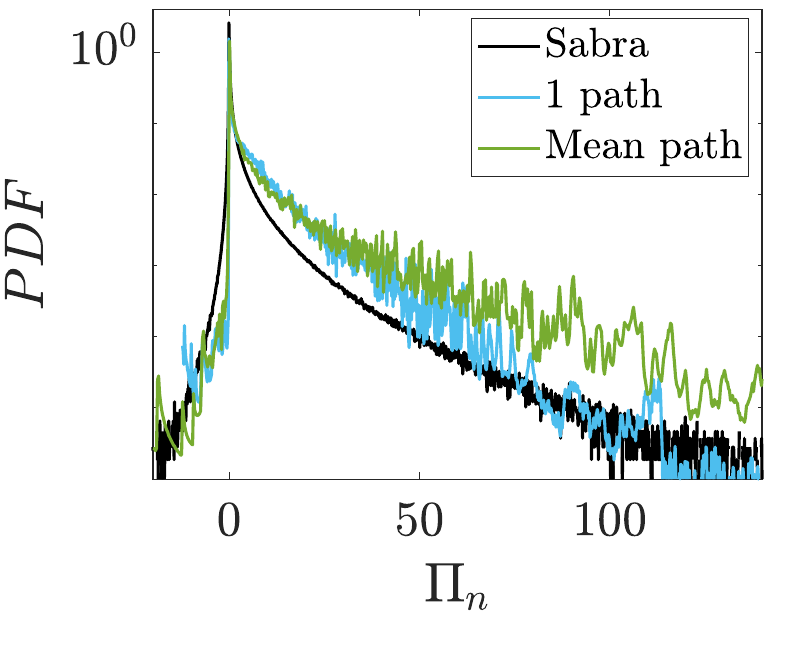}
  \label{fig:sindyflx8}
\end{subfigure}%
\begin{subfigure}{.333\textwidth}
  \centering
  \subcaption{$n=9$}
  \includegraphics[scale=0.41]{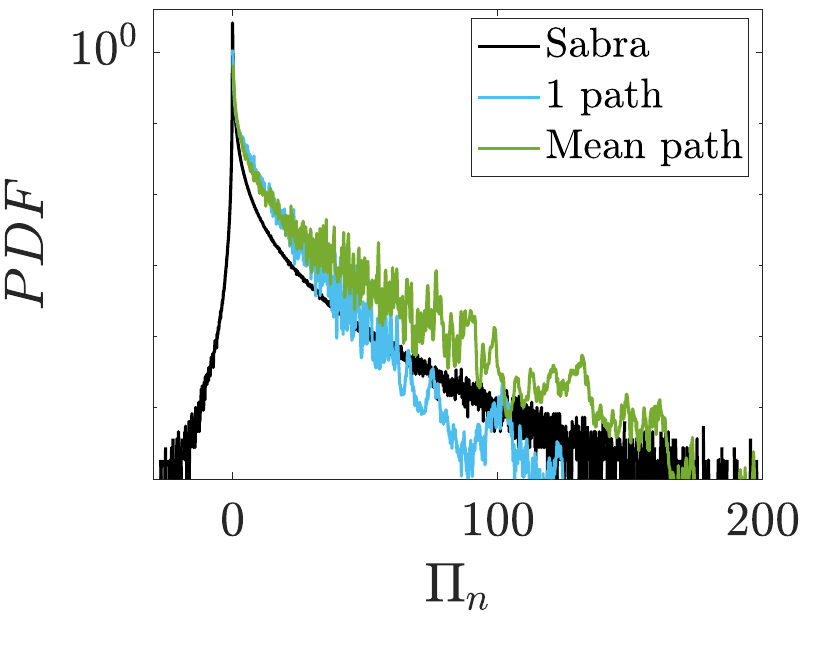}
  \label{fig:sindyflx9}
\end{subfigure}%
\caption{Cut-off shell is $s=9$. Performance of closures based on the stochastic version of the approximate dynamics found by SINDy, both using a single realization and taking a mean path. Comparison is made based on Sabra statistics of the fully resolved model. Figures \ref{fig:sindyre9}---\ref{fig:sindyre11} show normalized PDFs of real parts for the last shell of the reduced model and for the closure variables. Figures \ref{fig:sindyflx7}---\ref{fig:sindyflx9} show energy flux PDFs for the last shells of the reduced model. Figures \ref{fig:sindymjoint9} and \ref{fig:sindypjoint9} display moments of orders 2 up to 6 for reduced model simulations performed using one realization and a mean of several realizations respectively. Both are compared to Sabra moments in solid lines.}
\label{fig:sindys9}
\end{figure}

In figures \ref{fig:sindyre12}---\ref{fig:sindyre14} we display normalized PDFs of real parts for the cutoff shell $s=12$ and for  the closure variables. We can see that, unlike figures \ref{fig:re12}---\ref{fig:re14}, the closure PDFs do not display the tendency of becoming more and more heavy tailed. This is expected, given that these closures were built on data solely from the closure variables, and the approximate dynamic for their evolution is run completely decoupled from the reduced system. It is quite remarkable that, despite not including phases in its modeling, which is clear from the lack of back scattering in the energy flux PDF in figure \ref{fig:sindyflx12}, both \ref{fig:sindyflx10} and \ref{fig:sindyflx11} show much better agreement with the left legs of the PDFs than the VAE closures reported in figure \ref{fig:vae12}. In figures \ref{fig:sindymjoint} and \ref{fig:sindypjoint} we can see moments of orders 2 up to 6.

Figures \ref{fig:sindyre9}---\ref{fig:sindyre11} present normalized PDFs of real parts for the cutoff shell $s=9$ and the subsequent closure variables. In figures \ref{fig:sindyflx7}---\ref{fig:sindyflx9} we see energy flux PDFs for the last shells of the reduced model. Again, we point to the fact that this change in cutoff was performed only for the evolution of reduced model. The closure variables were a different realization of the stochastic process that solves \cref{eqn:stochsindy0,eqn:stochsindy1} for the One Path closure and the mean of $100$ different realizations for the Mean path closure. We see that qualitative behavior of these closures remained the same, and we remark that the oscillations we see in figure \ref{fig:sindymjoint} and \ref{fig:sindypjoint} are also present in \ref{fig:sindymjoint9} and \ref{fig:sindypjoint9}. Moreover, such oscillations seem to be at the same distance from the cutoff in both cases which indicates that these small discrepancies are due to proximity to the cutoff.

\begin{figure}[t!]
\begin{subfigure}{.333\textwidth}
  \centering
  \subcaption{$n=15$}
  \includegraphics[scale=0.42]{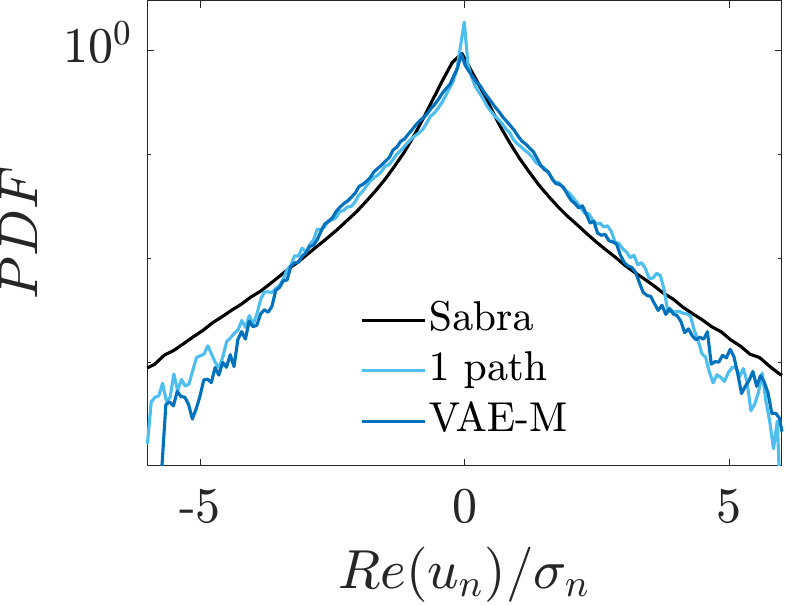}
  \label{fig:3sindyre15}
\end{subfigure}%
\begin{subfigure}{.333\textwidth}
  \centering
  \subcaption{$n=16$}
  \includegraphics[scale=0.42]{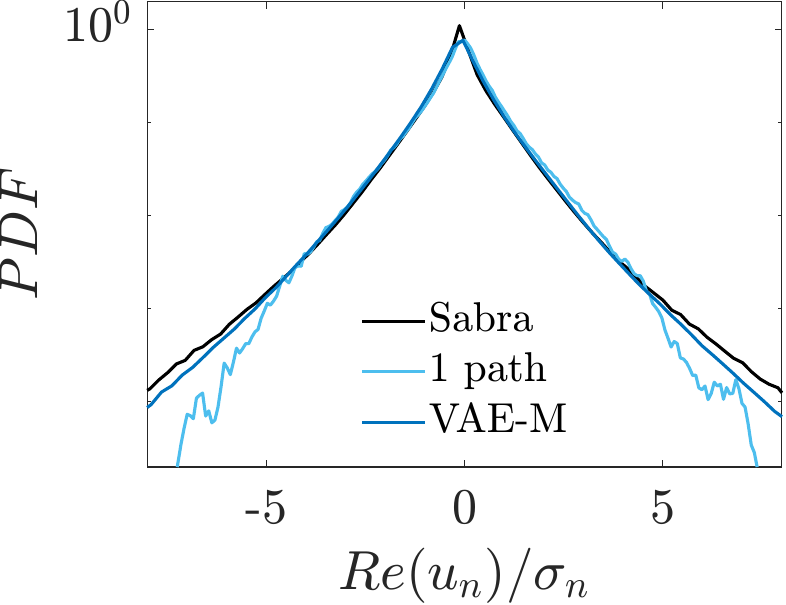}
  \label{fig:3sindyre16}
\end{subfigure}%
\begin{subfigure}{.333\textwidth}
  \centering
  \subcaption{$n=17$}
  \includegraphics[scale=0.42]{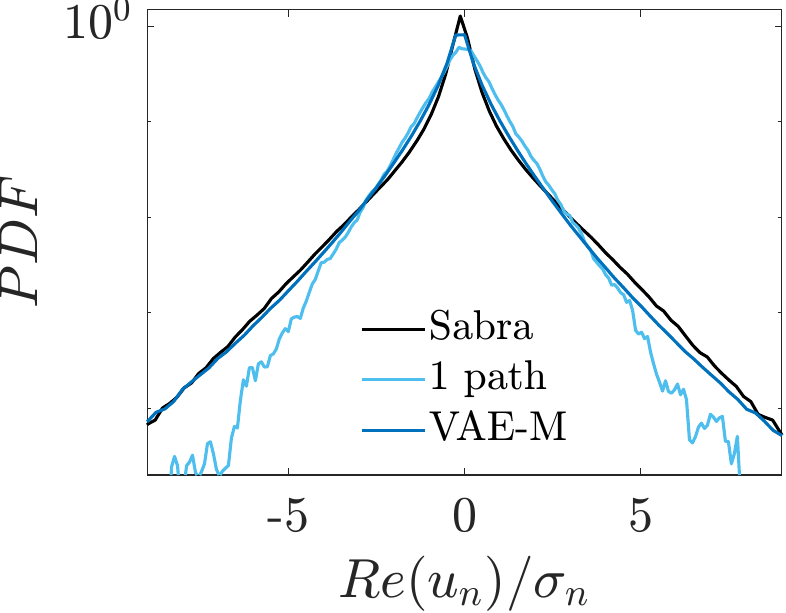}
  \label{fig:3sindyre17}
\end{subfigure}%

\begin{subfigure}{.5\textwidth}
  \centering
  \subcaption{One Path}
  \includegraphics[scale=0.6]{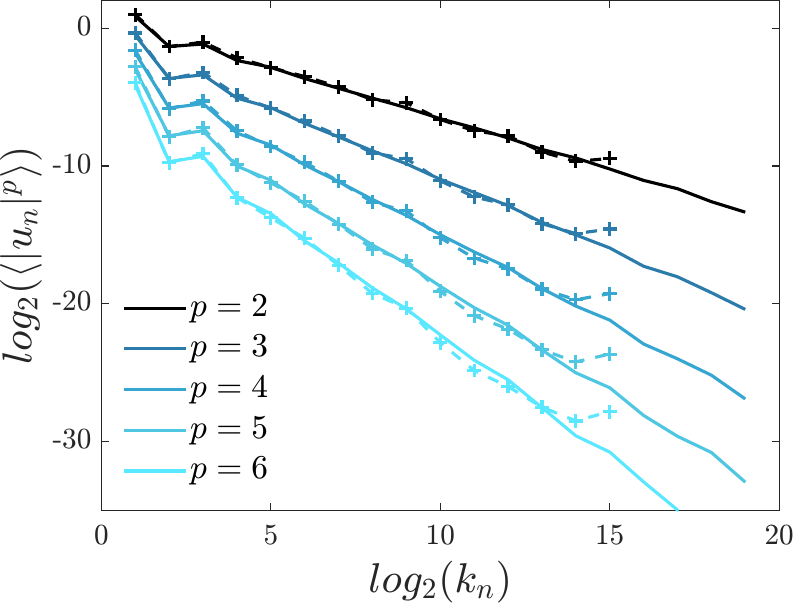}
  \label{fig:3sindymjoint}
\end{subfigure}%
\begin{subfigure}{.5\textwidth}
  \centering
  \subcaption{VAE-M}
  \includegraphics[scale=0.6]{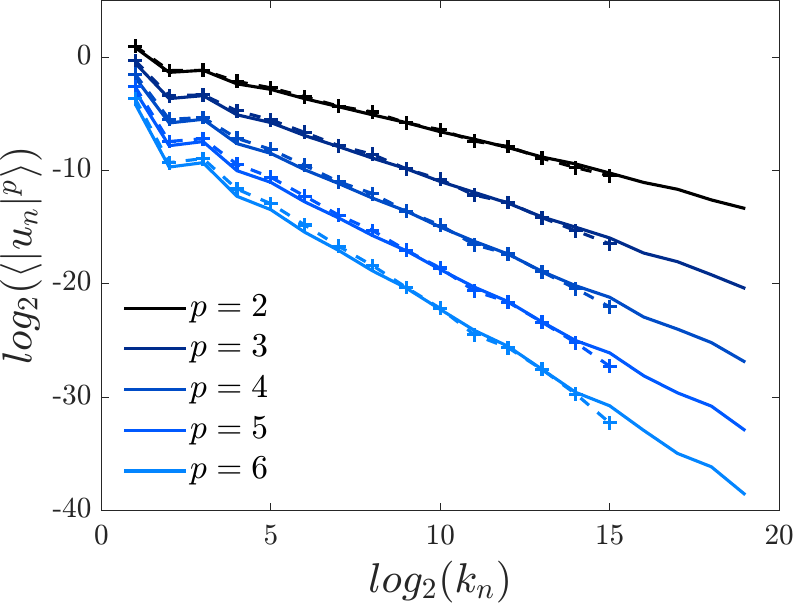}
  \label{fig:3vaemjoint}
\end{subfigure}%

\begin{subfigure}{.333\textwidth}
  \centering
  \subcaption{$n=13$}
  \includegraphics[scale=0.42]{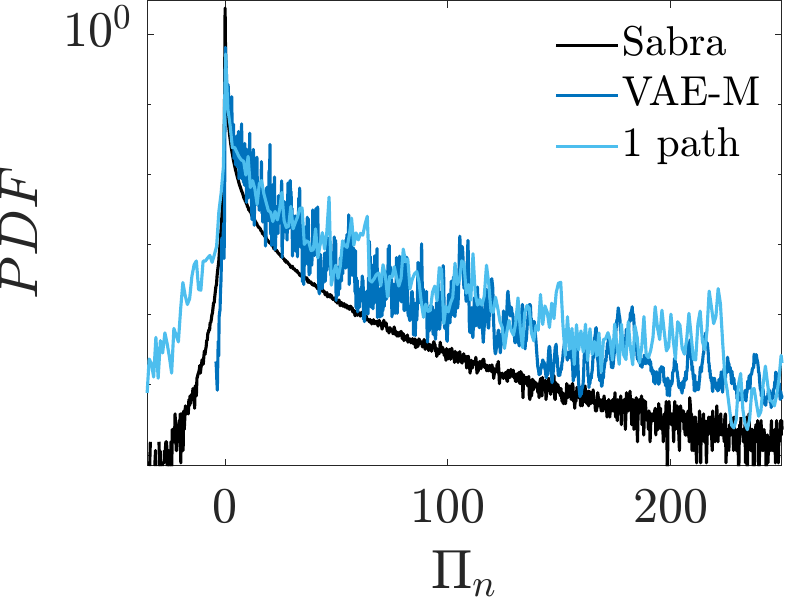}
  \label{fig:3sindyflx13}
\end{subfigure}%
\begin{subfigure}{.333\textwidth}
  \centering
  \subcaption{$n=14$}
  \includegraphics[scale=0.42]{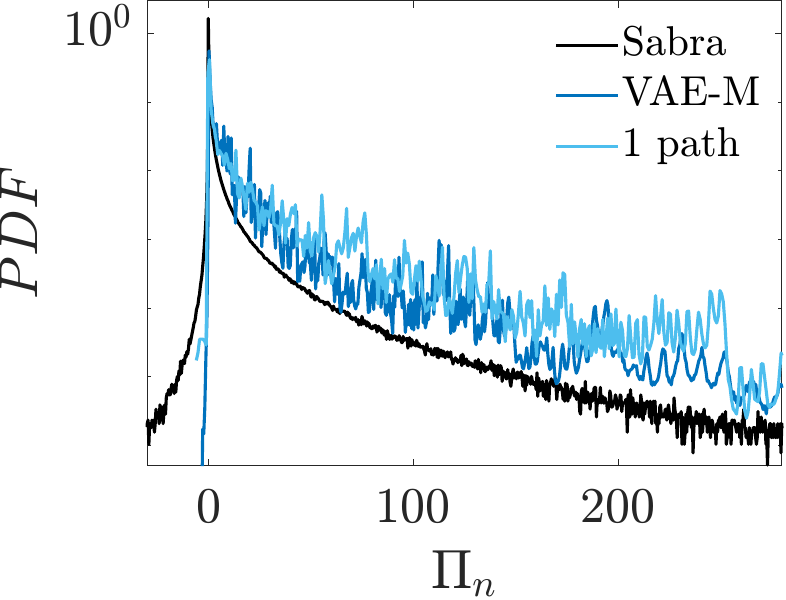}
  \label{fig:3sindyflx14}
\end{subfigure}%
\begin{subfigure}{.333\textwidth}
  \centering
  \subcaption{$n=15$}
  \includegraphics[scale=0.42]{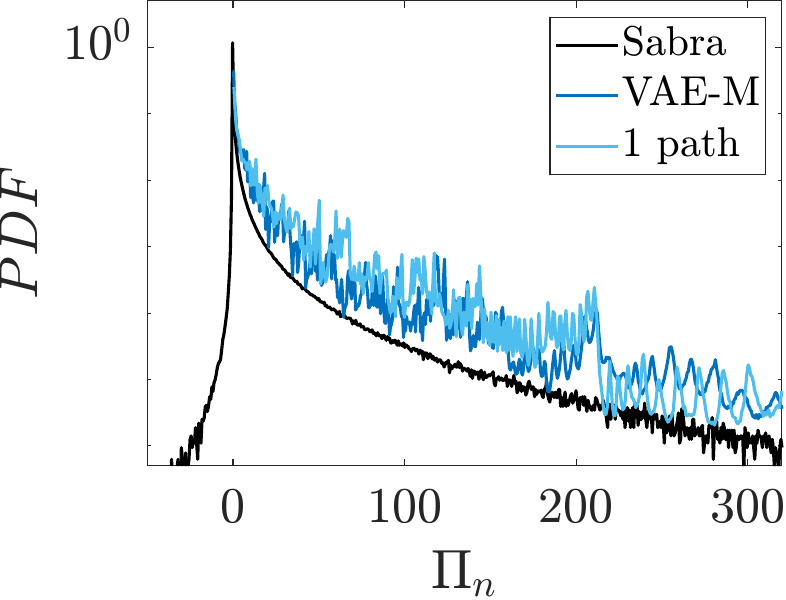}
  \label{fig:3sindyflx15}
\end{subfigure}%
\caption{Cut-off shell is $s=15$. Performance of closures based on the stochastic version of the approximate dynamics found by SINDy using a single realization and based on the trained VAE-M. Comparison is made based on Sabra statistics of the fully resolved model. Figures \ref{fig:sindyre12}---\ref{fig:sindyre14} show normalized PDFs of real parts for the last shell of the reduced model and for the closure variables. Figures \ref{fig:sindyflx9}---\ref{fig:sindyflx12} show energy flux PDFs for the last shells of the reduced model. Figures \ref{fig:sindymjoint} and \ref{fig:sindypjoint} display moments of orders 2 up to 6 for reduced model simulations performed using a single realization on the SINDy approximation and VAE-M respectively. Both are compared to Sabra moments in solid lines.}
\label{fig:sindys15}
\end{figure}

Now, we look at the performance of two closures, the One Path and the VAE-M, but for a cutoff $s=15$. In figures \ref{fig:3sindyre15}---\ref{fig:3sindyre17} we see normalized PDFs of real parts for the cutoff shell $s=15$ and the subsequent closure variables. We point that, at $n=18$, for our choices of parameters, Sabra is already approaching dissipation range. It is noteworthy that the VAE-M closure managed to recover quite well a statistic so deep in the inertial range as in figure \ref{fig:3sindyre17}. In figures \ref{fig:3sindyflx13}---\ref{fig:3sindyflx15} we see energy flux PDFs for the last shells of the reduced model.  In figures \ref{fig:3sindymjoint} and \ref{fig:3vaemjoint} the One Path closure and the VAE-M retained the same close-to-cutoff behavior of presenting a small discrepancy, always at the same distance from the cutoff. Again, we see that the qualitative behavior of these closures remained the same as in cutoffs $s=9$ and $s=12$. It is important to recall that all machine learning tools involved in these results were trained for $s=12$, which means that our reduced models display consistent behavior for scales both larger (smaller $s$) and smaller (larger cutoff $s$) than the one where they were trained.

\begin{figure}[t!]
\centering
\begin{subfigure}{.3\textwidth}
  \centering
  \subcaption{$s=9$}
  \includegraphics[scale=0.4]{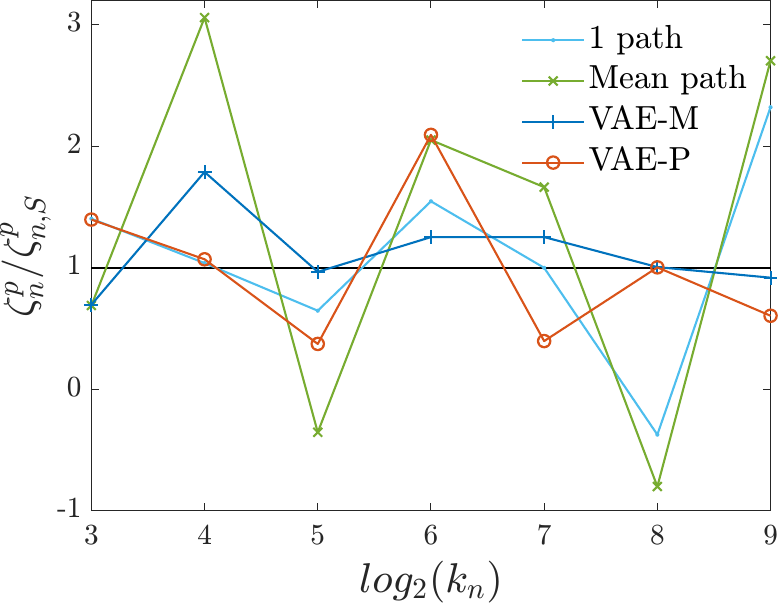}
  \label{fig:local9}
\end{subfigure}%
\begin{subfigure}{.3\textwidth}
  \centering
  \subcaption{$s=12$}
  \includegraphics[scale=0.4]{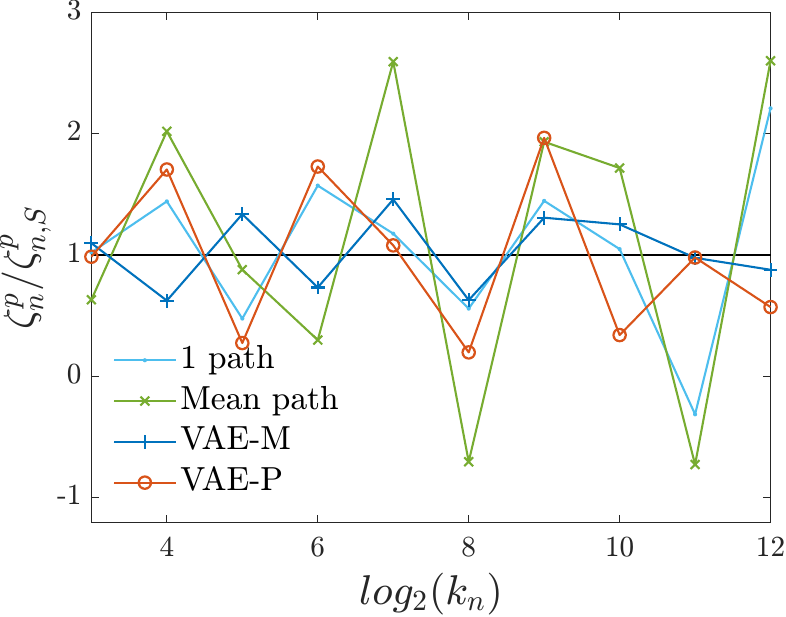}
  \label{fig:local12}
\end{subfigure}%
\begin{subfigure}{.3\textwidth}
  \centering
  \subcaption{$s=15$}
  \includegraphics[scale=0.4]{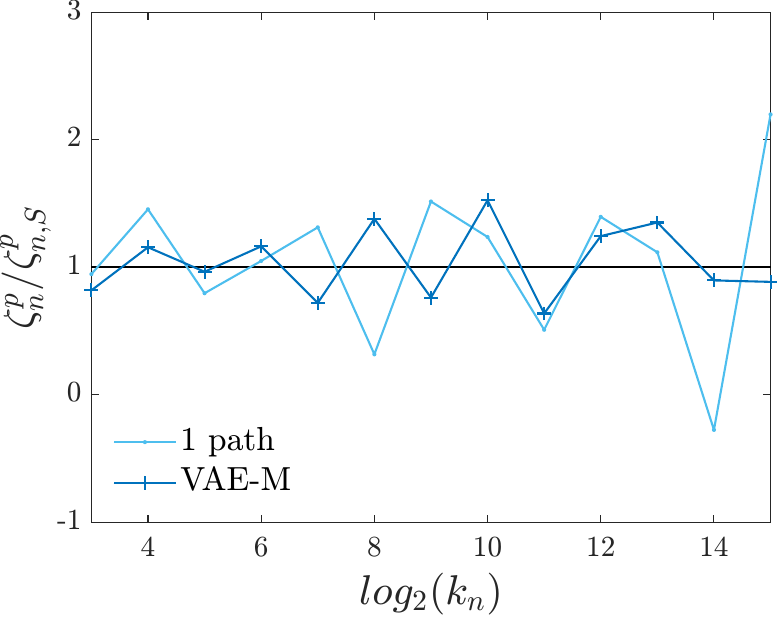}
  \label{fig:local15}
\end{subfigure}%
\caption{Local slopes for $p=2$ and different values of cutoff $s$, all normalized by the reference value $\zeta_{n,S}^p$.}
\label{fig:15}
\end{figure}

Lastly, in figure \ref{fig:15} we present the local slopes of second order moments, defined in equation \eqref{eqn:slopes}. We see that there is substantial oscillations around the reference value, which is expected from the general behavior of our models in moments recovery. We can also see some other instances of close-to-cutoff behavior discrepancies.

\section{Discussion} \label{sec:disc}

In a shell model, we want to find a way to resolve only large scales of motion and model small scales in a physically correct way. For Sabra, due to its particular non-linear coupling of scales, this reflects on writing expressions for the missing variables. Sabra statistics, however, are not universal throughout the inertial range due to intermittency, so we made use of a set of spatio-temporal scaling relations which encoded intermittency in the change of variables, therefore returning universal statistics and improving our chances of writing consistent closures. We then start searching for a suitable machine learning tool that can learn from data of a fully resolved Sabra simulation.

The VAEs, in general, show an improvement from previous work using the Hidden Symmetry framework, especially when we note that the PDFs of real parts displayed a gap that was not closed by any of the time-correlated closures written using Gaussian Mixtures Models, including the multi-time models reported in \cite{memyselfandi}. This is an important indicator that the robustness and suitability of the machine learning tool, employed either in the density estimation or in the dynamics recovery, does play an important role in the performance of the closures, as pointed in previous works \cite{euzinha}. 

From figure \ref{fig:vae12} paired with \ref{fig:vae9}, as well as figure \ref{fig:sindys} paired with \ref{fig:sindys9}, we see that the all closures performed consistently without the need to be retrained when we change the cut-off shells to a larger scale (smaller $s$) than the one where the tools were trained. Figure \ref{fig:sindys15} is an important result that points to the fact that the same machine learning tools can also be used to close the reduced systems at smaller scales (larger $s$). This is a feature inherited from the universality of statistics that we achieved due to the change of variables we applied earlier.

It is important to note that, when we fix phases at a strictly dissipating value one the VAE based closures, the structure functions in all cut-offs tend to stay below the Sabra ones, meaning such closures tend to slightly over-dissipate energy, which is visibly improved when we include phases in the modeling. This is also reflected on the PDFs for energy flux, where we can see that the inclusion of phases adds a significant contribution on the left leg of the PDF, which not only agrees more with the baseline Sabra statistics, but the closure that only includes modules can not recover.

It is vital that the VAEs perform a good estimation of the joint densities. In fact, our experiments indicate that whenever a VAE fails to produce a good estimation of the joint densities, the closure that makes use of such an approximation will not perform well, even if the component-wise PDFs of the generated data show good agreement with the training data. This would be a situation where figures \ref{fig:vaeU0} and \ref{fig:vaeU1} would show good agreement, but figures \ref{fig:jointtrue} and \ref{fig:jointvae} would look nothing alike.

On the other hand, the SINDy approach shows somewhat surprising results, given that the quality of the approximation of the closure variables has not been evaluated regarding how well the densities are being approximated, but instead from the qualitative behavior of two qualitatively different systems. Still, from the phase portrait in figure \ref{fig:portrait}, we see that the dynamic recovered by SINDy, which we later made stochastic by adding a diffusion term, occupies a region in phase space that significantly overlaps with the one occupied by the rescaled Sabra variables.

The many trajectories we computed behaved, in their vast majority, much like the one displayed in figure \ref{fig:portrait}, starting at the same initial condition, approaching the origin and then quickly escaping towards the limit-cycle-like region we see in the figure. Out of 100 trajectories we generated and inspected, four of them initially approached the origin and remained there for an extended period of time, later (at around $\tau = 4000$) escaping towards the same region as the others. One single trajectory approached the origin and remained there for all the time it was numerically evolved. 

Both SINDy based closures performed similarly well, with the Mean-Path struggling slightly more to recover moments. This is not surprising, given that taking ensemble averages as we are doing here greatly homogenizes what once was a much more detailed trajectory. It is important to note that the SINDy closures are also improving on the GMM results previously reported, more visibly so on moments and energy flux PDFs.

In general, what we see in these results is a confirmation of the projection that, on the Hidden-Symmetry framework, more powerful machine learning tools will improve on the accuracy of statistics recovery. In this sense, other works \cite{sabrann,sabrann25,lstmsabra}, where the authors train their machine learning tools directly on Sabra statistics, present much more accurate closures than ours, recovering statistics and local slopes with remarkable precision. More recently \cite{sabrann25} it was pointed out that such closures may face a loss of scale invariance on the multipliers statistics near the cutoff. This makes the universality of statistics found in the Hidden-Symmetry framework much more attractive as a potential application of more robust machine learning tools to closure problems in shell models. We have seen throughout this report that the closures trained on the Hidden-Symmetry framework are easily generalizable to different cutoffs, and the employment of more suitable tools presented significant improvement from earlier results in the same framework.

However, while our results improve upon previous work within the same framework, further enhancements are possible, for instance, by incorporating phases into the SINDy approach (forcing properties such as physical periodicity). This was not achieved here due to the periodic nature of phases, which introduces jump-like discontinuities in their time evolution, complicating their inclusion. One can also investigate further different choices of $\sigma$ for the diffusion term in the stochastic SINDy approach. Our results could also be majorly improved by including time conditioning, either in an alternative type of variational neural network, or by training a conditional VAE. 

We report these results in the hope that it contributes to our understanding not only of closure problems as a whole, but also in better comprehending the Hidden Symmetry framework and its features. The expectation that these results could be useful in writing closures for Navier-Stokes equations is an important and motivating factor of the choices we made regarding probabilistic closures that work cutoff-independently.


\section*{Acknowledgements}
This work was supported by the Serrapilheira Institute (grant number Serra – R-2111-39718)

\section*{Conflict of Interest}

The authors have no conflicts to disclose.

\section*{Data availability}

The data generated and analyzed in this study is available from the corresponding author on reasonable request.

\bibliography{sample}

\begin{thebibliography}{54}
\providecommand{\natexlab}[1]{#1}
\providecommand{\url}[1]{\texttt{#1}}
\expandafter\ifx\csname urlstyle\endcsname\relax
  \providecommand{\doi}[1]{doi: #1}\else
  \providecommand{\doi}{doi: \begingroup \urlstyle{rm}\Url}\fi

\bibitem[Li et~al.(2009)Li, Chevillard, Eyink, and Meneveau]{clos_eyink}
Yi~Li, Laurent Chevillard, Gregory Eyink, and Charles Meneveau.
\newblock Matrix exponential-based closures for the turbulent subgrid-scale stress tensor.
\newblock \emph{Phys. Rev. E, Statistical, nonlinear, and soft matter physics}, 79:\penalty0 016305, 02 2009.
\newblock \doi{10.1103/PhysRevE.79.016305}.

\bibitem[Frisch(1995)]{frisch}
Uriel Frisch.
\newblock \emph{Turbulence: The Legacy of A.N. Kolmogorov}.
\newblock 11 1995.
\newblock ISBN 9780521451031.
\newblock \doi{10.1017/CBO9781139170666}.

\bibitem[Reynolds(1895)]{reynoldsaverage}
Osborne Reynolds.
\newblock On the dynamical theory of incompressible viscous fluids and the determination of the criterion.
\newblock \emph{Philosophical Transactions of the Royal Society of London A}, \penalty0 (186):\penalty0 123--164, 1895.
\newblock URL \url{http://www.jstor.org/stable/109431}.

\bibitem[Smagorinsky(1963)]{les_1}
J.~Smagorinsky.
\newblock General circulation experiments with the primitive equations: I. the basic experiment.
\newblock \emph{Monthly Weather Review}, 91\penalty0 (3):\penalty0 99 -- 164, 1963.
\newblock \doi{10.1175/1520-0493(1963)091<0099:GCEWTP>2.3.CO;2}.

\bibitem[Pope(2000)]{pope2000turbulent}
Stephen~B. Pope.
\newblock \emph{Turbulent Flows}.
\newblock Cambridge University Press, 2000.
\newblock \doi{10.1017/CBO9781316179475}.

\bibitem[Chen et~al.(2023)Chen, Farhat, and Lunasin]{nudging}
Nan Chen, Aseel Farhat, and Evelyn Lunasin.
\newblock Data assimilation with model error: Analytical and computational study for sabra shell model.
\newblock \emph{Physica D: Nonlinear Phenomena}, 443:\penalty0 133552, 2023.
\newblock ISSN 0167-2789.
\newblock \doi{https://doi.org/10.1016/j.physd.2022.133552}.
\newblock URL \url{https://www.sciencedirect.com/science/article/pii/S0167278922002561}.

\bibitem[Alexander et~al.(2008)Alexander, Johnson, Eyink, and Kevrekidis]{clos_eqfree}
Francis~J. Alexander, Gregory Johnson, Gregory~L. Eyink, and Ioannis~G. Kevrekidis.
\newblock Equation-free implementation of statistical moment closures.
\newblock \emph{Phys. Rev. E}, 77:\penalty0 026701, Feb 2008.
\newblock \doi{10.1103/PhysRevE.77.026701}.
\newblock URL \url{https://link.aps.org/doi/10.1103/PhysRevE.77.026701}.

\bibitem[Lorenz(1969)]{Lorenz69}
Edward~Norton Lorenz.
\newblock The predictability of a flow which possesses many scales of motion.
\newblock \emph{Tellus A}, 21:\penalty0 289--307, 1969.

\bibitem[Leith and Kraichnan(1972)]{leithkraich}
C.~E. Leith and R.~H. Kraichnan.
\newblock Predictability of turbulent flows.
\newblock \emph{Journal of Atmospheric Sciences}, 29\penalty0 (6):\penalty0 1041 -- 1058, 1972.
\newblock \doi{10.1175/1520-0469(1972)029<1041:POTF>2.0.CO;2}.

\bibitem[Ruelle(1979)]{ruelle}
David Ruelle.
\newblock Microscopic fluctuations and turbulence.
\newblock \emph{Physics Letters A}, 72\penalty0 (2):\penalty0 81--82, 1979.
\newblock ISSN 0375-9601.
\newblock \doi{https://doi.org/10.1016/0375-9601(79)90653-4}.
\newblock URL \url{https://www.sciencedirect.com/science/article/pii/0375960179906534}.

\bibitem[Eyink(1996)]{eyinknoise}
Gregory~L. Eyink.
\newblock Turbulence noise.
\newblock \emph{Journal of Statistical Physics}, 83:\penalty0 81--82, 1996.
\newblock ISSN 955-1019.
\newblock \doi{https://doi.org/10.1007/BF02179551}.

\bibitem[Boffetta et~al.(2002)Boffetta, Cencini, Falcioni, and Vulpiani]{massimum}
G.~Boffetta, M.~Cencini, M.~Falcioni, and A.~Vulpiani.
\newblock Predictability: a way to characterize complexity.
\newblock \emph{Physics Reports}, 356\penalty0 (6):\penalty0 367--474, 2002.
\newblock ISSN 0370-1573.
\newblock \doi{https://doi.org/10.1016/S0370-1573(01)00025-4}.
\newblock URL \url{https://www.sciencedirect.com/science/article/pii/S0370157301000254}.

\bibitem[Palmer et~al.(2014)Palmer, Döring, and Seregin]{Palmer}
T~N Palmer, A~Döring, and G~Seregin.
\newblock The real butterfly effect.
\newblock \emph{Nonlinearity}, 27\penalty0 (9):\penalty0 R123--R141, aug 2014.
\newblock \doi{10.1088/0951-7715/27/9/r123}.
\newblock URL \url{https://doi.org/10.1088/0951-7715/27/9/r123}.

\bibitem[Mailybaev(2016)]{Mailybaev_2016}
Alexei~A Mailybaev.
\newblock Spontaneously stochastic solutions in one-dimensional inviscid systems.
\newblock \emph{Nonlinearity}, 29\penalty0 (8):\penalty0 2238--2252, jun 2016.
\newblock \doi{10.1088/0951-7715/29/8/2238}.
\newblock URL \url{https://doi.org/10.1088/0951-7715/29/8/2238}.

\bibitem[Mailybaev(2017)]{Mailybaev_2017}
Alexei~A Mailybaev.
\newblock Toward analytic theory of the rayleigh{\textendash}taylor instability: lessons from a toy model.
\newblock \emph{Nonlinearity}, 30\penalty0 (6):\penalty0 2466--2484, may 2017.
\newblock \doi{10.1088/1361-6544/aa6eb5}.
\newblock URL \url{https://doi.org/10.1088/1361-6544/aa6eb5}.

\bibitem[Biferale et~al.(2018)Biferale, Boffetta, Mailybaev, and Scagliarini]{RTbifbof}
L.~Biferale, G.~Boffetta, A.~A. Mailybaev, and A.~Scagliarini.
\newblock Rayleigh-taylor turbulence with singular nonuniform initial conditions.
\newblock \emph{Phys. Rev. Fluids}, 3:\penalty0 092601, Sep 2018.
\newblock \doi{10.1103/PhysRevFluids.3.092601}.
\newblock URL \url{https://link.aps.org/doi/10.1103/PhysRevFluids.3.092601}.

\bibitem[Thalabard et~al.(2020)Thalabard, Bec, and Mailybaev]{spont1}
Simon Thalabard, Jeremie Bec, and Alexei Mailybaev.
\newblock From the butterfly effect to spontaneous stochasticity in singular shear flows.
\newblock \emph{Communications Physics}, 3, 07 2020.
\newblock \doi{10.1038/s42005-020-0391-6}.

\bibitem[Bandak et~al.(2024)Bandak, Mailybaev, Eyink, and Goldenfeld]{bandak2024}
Dmytro Bandak, Alexei~A. Mailybaev, Gregory~L. Eyink, and Nigel Goldenfeld.
\newblock Spontaneous stochasticity amplifies even thermal noise to the largest scales of turbulence in a few eddy turnover times.
\newblock \emph{Phys. Rev. Lett.}, 132:\penalty0 104002, Mar 2024.
\newblock \doi{10.1103/PhysRevLett.132.104002}.
\newblock URL \url{https://link.aps.org/doi/10.1103/PhysRevLett.132.104002}.

\bibitem[Gledzer(1973)]{gled}
E.~B. Gledzer.
\newblock System of hydrodynamic type admitting two quadratic integrals of motion.
\newblock \emph{Soviet Physics Doklady}, 18:\penalty0 216, oct 1973.

\bibitem[Yamada and Ohkitani(1988)]{goy}
Michio Yamada and Koji Ohkitani.
\newblock Lyapunov spectrum of a model of two-dimensional turbulence.
\newblock \emph{Phys. Rev. Lett.}, 60:\penalty0 983--986, Mar 1988.
\newblock \doi{10.1103/PhysRevLett.60.983}.
\newblock URL \url{https://link.aps.org/doi/10.1103/PhysRevLett.60.983}.

\bibitem[Biferale(2003)]{lucashell}
Luca Biferale.
\newblock Shell models of energy cascade in turbulence.
\newblock \emph{Annual Review of Fluid Mechanics}, 35\penalty0 (1):\penalty0 441--468, 2003.
\newblock \doi{10.1146/annurev.fluid.35.101101.161122}.

\bibitem[L’vov et~al.(1998)L’vov, Podivilov, Pomyalov, Procaccia, and Vandembroucq]{lvov}
Victor~S. L’vov, Evgenii Podivilov, Anna Pomyalov, Itamar Procaccia, and Damien Vandembroucq.
\newblock Improved shell model of turbulence.
\newblock \emph{Phys. Rev. E}, 58\penalty0 (2):\penalty0 1811--1822, 1998.

\bibitem[Benzi et~al.(1993)Benzi, Biferale, and Parisi]{benzi}
R.~Benzi, L.~Biferale, and G.~Parisi.
\newblock On intermittency in a cascade model for turbulence.
\newblock \emph{Physica D: Nonlinear Phenomena}, 65\penalty0 (1):\penalty0 163--171, 1993.
\newblock ISSN 0167-2789.
\newblock \doi{https://doi.org/10.1016/0167-2789(93)90012-P}.
\newblock URL \url{https://www.sciencedirect.com/science/article/pii/016727899390012P}.

\bibitem[Eyink et~al.(2003)Eyink, Chen, and Chen]{eyink2003}
Gregory Eyink, Shiyi Chen, and Qiaoning Chen.
\newblock Gibbsian hypothesis in turbulence.
\newblock \emph{Journal of Statistical Physics}, 113:\penalty0 719--740, 12 2003.
\newblock \doi{10.1023/A:1027304501435}.

\bibitem[Kolmogorov(1962)]{kolmogorov_1962}
A.~N. Kolmogorov.
\newblock A refinement of previous hypotheses concerning the local structure of turbulence in a viscous incompressible fluid at high reynolds number.
\newblock \emph{Journal of Fluid Mechanics}, 13\penalty0 (1):\penalty0 82–85, 1962.
\newblock \doi{10.1017/S0022112062000518}.

\bibitem[Biferale et~al.(2017)Biferale, Mailybaev, and Parisi]{Biferale_2017}
Luca Biferale, Alexei~A. Mailybaev, and Giorgio Parisi.
\newblock Optimal subgrid scheme for shell models of turbulence.
\newblock \emph{Phys. Rev. E}, 95\penalty0 (4), apr 2017.
\newblock \doi{10.1103/physreve.95.043108}.
\newblock URL \url{https://doi.org/10.1103%2Fphysreve.95.043108}.

\bibitem[Mailybaev(2021)]{alexei}
Alexei~A. Mailybaev.
\newblock Hidden scale invariance of intermittent turbulence in a shell model.
\newblock \emph{Phys. Rev. Fluids}, 6:\penalty0 L012601, Jan 2021.
\newblock \doi{10.1103/PhysRevFluids.6.L012601}.
\newblock URL \url{https://link.aps.org/doi/10.1103/PhysRevFluids.6.L012601}.

\bibitem[Mailybaev and Thalabard(2022)]{Mailybaev2022HiddenSI}
Alexei~A Mailybaev and Simon Thalabard.
\newblock Hidden scale invariance in navier--stokes intermittency.
\newblock \emph{Philosophical Transactions of the Royal Society A}, 380\penalty0 (2218):\penalty0 20210098, 2022.

\bibitem[Domingues~Lemos and Mailybaev(2024)]{euzinha}
J.~Domingues~Lemos and A.~A. Mailybaev.
\newblock Data-based approach for time-correlated closures of turbulence models.
\newblock \emph{Phys. Rev. E}, 109:\penalty0 025101, Feb 2024.
\newblock \doi{10.1103/PhysRevE.109.025101}.
\newblock URL \url{https://link.aps.org/doi/10.1103/PhysRevE.109.025101}.

\bibitem[Bishop(2006)]{bishop}
Christopher~M. Bishop.
\newblock \emph{Pattern Recognition and Machine Learning}.
\newblock Information science and statistics. Springer, 1st ed. 2006. corr. 2nd printing edition, 2006.
\newblock ISBN 9780387310732,0387310738.

\bibitem[Ortali et~al.(2022)Ortali, Corbetta, Rozza, and Toschi]{lstmsabra}
Giulio Ortali, Alessandro Corbetta, Gianluigi Rozza, and Federico Toschi.
\newblock Numerical proof of shell model turbulence closure.
\newblock \emph{Phys. Rev. Fluids}, 7:\penalty0 L082401, Aug 2022.
\newblock \doi{10.1103/PhysRevFluids.7.L082401}.
\newblock URL \url{https://link.aps.org/doi/10.1103/PhysRevFluids.7.L082401}.

\bibitem[Freitas et~al.(2024)Freitas, Um, Desbrun, Buzzicotti, and Biferale]{sabrann}
André Freitas, Kiwon Um, Mathieu Desbrun, Michele Buzzicotti, and Luca Biferale.
\newblock Solver-in-the-loop approach to turbulence closure, 2024.
\newblock URL \url{https://arxiv.org/abs/2411.13194}.

\bibitem[Kingma(2013)]{vae}
Diederik~P Kingma.
\newblock Auto-encoding variational bayes.
\newblock \emph{arXiv preprint arXiv:1312.6114}, 2013.

\bibitem[Brunton et~al.(2016)Brunton, Proctor, and Kutz]{sindy}
Steven~L. Brunton, Joshua~L. Proctor, and J.~Nathan Kutz.
\newblock Discovering governing equations from data by sparse identification of nonlinear dynamical systems.
\newblock \emph{Proceedings of the National Academy of Sciences}, 113\penalty0 (15):\penalty0 3932--3937, 2016.
\newblock \doi{10.1073/pnas.1517384113}.
\newblock URL \url{https://www.pnas.org/doi/abs/10.1073/pnas.1517384113}.

\bibitem[Frisch and Vergassola(1991)]{frisch1991prediction}
U~Frisch and M~Vergassola.
\newblock A prediction of the multifractal model: the intermediate dissipation range.
\newblock \emph{Europhysics Letters}, 14\penalty0 (5):\penalty0 439, 1991.

\bibitem[Mailybaev(2023)]{mailybaev2023hidden}
Alexei~A Mailybaev.
\newblock Hidden scale invariance of turbulence in a shell model: From forcing to dissipation scales.
\newblock \emph{Phys. Rev. Fluids}, 8\penalty0 (5):\penalty0 054605, 2023.

\bibitem[Mailybaev(2022)]{mailybaev2022hidden}
Alexei~A Mailybaev.
\newblock Hidden spatiotemporal symmetries and intermittency in turbulence.
\newblock \emph{Nonlinearity}, 35\penalty0 (7):\penalty0 3630, 2022.

\bibitem[Ramachandram and Taylor(2017)]{nonvae}
Dhanesh Ramachandram and Graham~W. Taylor.
\newblock Deep multimodal learning: A survey on recent advances and trends.
\newblock \emph{IEEE Signal Processing Magazine}, 34\penalty0 (6):\penalty0 96--108, 2017.
\newblock \doi{10.1109/MSP.2017.2738401}.

\bibitem[Schmidhuber(2015)]{reviewvae}
Jürgen Schmidhuber.
\newblock Deep learning in neural networks: An overview.
\newblock \emph{Neural Networks}, 61:\penalty0 85--117, 2015.
\newblock ISSN 0893-6080.
\newblock \doi{https://doi.org/10.1016/j.neunet.2014.09.003}.
\newblock URL \url{https://www.sciencedirect.com/science/article/pii/S0893608014002135}.

\bibitem[Belov and Armstrong(2011)]{divkl}
Dmitry Belov and Ronald Armstrong.
\newblock Distributions of the kullback-leibler divergence with applications.
\newblock \emph{The British journal of mathematical and statistical psychology}, 64:\penalty0 291--309, 05 2011.
\newblock \doi{10.1348/000711010X522227}.

\bibitem[Zhang and Schaeffer(2019)]{conv}
Linan Zhang and Hayden Schaeffer.
\newblock On the convergence of the sindy algorithm.
\newblock \emph{Multiscale Modeling \& Simulation}, 17\penalty0 (3):\penalty0 948--972, 2019.
\newblock \doi{10.1137/18M1189828}.
\newblock URL \url{https://doi.org/10.1137/18M1189828}.

\bibitem[Wentz and Doostan(2023)]{noisysindy}
Jacqueline Wentz and Alireza Doostan.
\newblock Derivative-based sindy (dsindy): Addressing the challenge of discovering governing equations from noisy data.
\newblock \emph{Computer Methods in Applied Mechanics and Engineering}, 413:\penalty0 116096, 2023.
\newblock ISSN 0045-7825.
\newblock \doi{https://doi.org/10.1016/j.cma.2023.116096}.
\newblock URL \url{https://www.sciencedirect.com/science/article/pii/S0045782523002207}.

\bibitem[Fiorini et~al.(2024)Fiorini, Flint, Fostier, Franck, Hashemi, Michel-Dansac, and Tenachi]{halsindy}
Camilla Fiorini, Cl{\'e}ment Flint, Louis Fostier, Emmanuel Franck, Reyhaneh Hashemi, Victor Michel-Dansac, and Wassim Tenachi.
\newblock {Generalizing the SINDy approach with nested neural networks}.
\newblock working paper or preprint, April 2024.
\newblock URL \url{https://hal.science/hal-04557263}.

\bibitem[Fasel et~al.(2022)Fasel, Kutz, Brunton, and Brunton]{esindy}
Urban Fasel, J.~Kutz, Bingni Brunton, and Steven Brunton.
\newblock Ensemble-sindy: Robust sparse model discovery in the low-data, high-noise limit, with active learning and control.
\newblock \emph{Proceedings of the Royal Society A: Mathematical, Physical and Engineering Sciences}, 478, 04 2022.
\newblock \doi{10.1098/rspa.2021.0904}.

\bibitem[Breiman(1996)]{Breiman1996BaggingP}
L.~Breiman.
\newblock Bagging predictors.
\newblock \emph{Machine Learning}, 24:\penalty0 123--140, 1996.
\newblock URL \url{https://api.semanticscholar.org/CorpusID:47328136}.

\bibitem[{\O}ksendal(2014)]{oksendal}
Bernt {\O}ksendal.
\newblock \emph{{Stochastic Differential Equations: An Introduction with Applications (Universitext)}}.
\newblock Springer, 6th edition, January 2014.
\newblock ISBN 3540047581.
\newblock URL \url{http://www.amazon.com/exec/obidos/redirect?tag=citeulike07-20\&path=ASIN/3540047581}.

\bibitem[Mandal and Tiwari(2024)]{Mandal2024}
Sayan Mandal and Pankaj~Kumar Tiwari.
\newblock Schooling behavior in a generalist predator–prey system: exploring fear, refuge and cooperative strategies in a stochastic environment.
\newblock \emph{The European Physical Journal Plus}, 139\penalty0 (1):\penalty0 29, 2024.
\newblock ISSN 2190-5444.
\newblock \doi{10.1140/epjp/s13360-023-04787-4}.
\newblock URL \url{https://doi.org/10.1140/epjp/s13360-023-04787-4}.

\bibitem[Kloeden(1999)]{eulermaru}
Peter~E Kloeden.
\newblock \emph{Numerical solution of stochatic differential equations}.
\newblock Applications of mathematics ; 23. Springer, Berlin [etc.], corr. 3rd printing. edition, 1999.
\newblock ISBN 3-540-54062-8.

\bibitem[Freitas et~al.(2025)Freitas, Um, Desbrun, Buzzicotti, and Biferale]{sabrann25}
André Freitas, Kiwon Um, Mathieu Desbrun, Michele Buzzicotti, and Luca Biferale.
\newblock A posteriori closure of turbulence models: are symmetries preserved ?, 2025.
\newblock URL \url{https://arxiv.org/abs/2504.03870}.

\bibitem[Householder(1941)]{Householder1941relu}
Alston~S. Householder.
\newblock A theory of steady-state activity in nerve-fiber networks: I. definitions and preliminary lemmas.
\newblock \emph{The Bulletin of Mathematical Biophysics}, 3\penalty0 (2):\penalty0 63--69, 1941.
\newblock ISSN 1522-9602.
\newblock \doi{10.1007/BF02478220}.
\newblock URL \url{https://doi.org/10.1007/BF02478220}.

\bibitem[He et~al.(2015)He, Zhang, Ren, and Sun]{heinit}
Kaiming He, Xiangyu Zhang, Shaoqing Ren, and Jian Sun.
\newblock Delving deep into rectifiers: Surpassing human-level performance on imagenet classification.
\newblock In \emph{2015 IEEE International Conference on Computer Vision (ICCV)}, pages 1026--1034, 2015.
\newblock \doi{10.1109/ICCV.2015.123}.

\bibitem[Kingma and Ba(2017)]{adam}
Diederik~P. Kingma and Jimmy Ba.
\newblock Adam: A method for stochastic optimization, 2017.
\newblock URL \url{https://arxiv.org/abs/1412.6980}.

\bibitem[Coddington and Levinson(1955)]{poinc}
A.~Coddington and N.~Levinson.
\newblock \emph{Theory of Ordinary Differential Equations}.
\newblock International series in pure and applied mathematics. McGraw-Hill, 1955.
\newblock ISBN 9780070992566.
\newblock URL \url{https://books.google.com.br/books?id=LvNQAAAAMAAJ}.

\bibitem[Lemos(2022)]{memyselfandi}
J.~Domingues Lemos.
\newblock \emph{Data-based approach for time-correlated closures of turbulence models}.
\newblock PhD thesis, National Institute for Pure and Applied Mathematics, 2022.
\newblock URL \url{https://impa.br/wp-content/uploads/2022/11/dout_tese_Julia_Domingues_Lemos.pdf}.

\end{thebibliography}

\end{document}